\PassOptionsToPackage{english}{babel}
\RequirePackage{lineno}
\documentclass[preprint,eqsecnum,aps, nofootinbib,12pt]{revtex4}
\usepackage{subfigure}
\usepackage[geometry]{ifsym}

\usepackage[english]{babel}
\usepackage{setspace} 
\usepackage{url}
\usepackage{graphicx}
\usepackage{amssymb} %maths
\usepackage{amsmath} %maths
\usepackage{amsfonts}
\usepackage{placeins}
\usepackage[multiple]{footmisc}
\usepackage[font=small,labelfont=bf]{caption}
\usepackage{color}
\usepackage[usenames,dvipsnames]{xcolor}

\definecolor{Crimson}{rgb}{0.6471, 0.1098, 0.1882}
\definecolor{Gold}{rgb}{0.87890625, 0.68359375, 0.14453125}
\definecolor{SeaGreen}{rgb}{0.421875, 0.70703125, 0.609375}
\definecolor{Grey}{rgb}{0.42578125, 0.4296875, 0.44140625}

\definecolor{colourOfFootnotes}{HTML}{B40404}
\definecolor{colourOfCitations}{HTML}{B40404}
\definecolor{colouroflinks}{rgb}{0.3,0.0,0.0}
\definecolor{charcoal-gray}{gray}{0.3}

\usepackage[pdfauthor={Benjamin Stephen MacPerson Barlow}, colorlinks=true, linkcolor= colouroflinks, citecolor= colourOfCitations, unicode=true, bookmarks=true, bookmarksnumbered=true]{hyperref}
\usepackage{verbatim}

\newcommand{\alphaEntropic}{\alpha_{\emph{entropic}}}
\newcommand{\alphaDirect}{\alpha_{\emph{direct}}}
\newcommand{\alphaTriangulated}{\alpha^{\text{tri}}}

\newcommand{\K}{\ensuremath{K_{\!A}}}
\newcommand{\F}{\ensuremath{F_{\!z}}}
\newcommand{\kBoltzmann}{k_{B}}

\newcommand{\Mell}{_{\text{Mell}}}
\newcommand{\HS}{_{\text{HS}}}

\newcommand{\VSS}{V_{SS}}
\newcommand{\VLJ}{V_{LJ}}

\renewcommand{\d}{d}
\newcommand{\nano}{\text{n}}
\newcommand{\pico}{\text{p}}

\newcommand{\Newton}{\text{N}}
\newcommand{\Joule}{\text{J}}
\newcommand{\metre}{\text{m}}
\newcommand{\second}{\text{s}}
\newcommand{\gram}{\text{g}}
\newcommand{\kilo}{\text{k}}
\newcommand{\Mass}[0]{\scalebox{0.72}{\ensuremath{\mathbf{M}}}} 
\newcommand{\Length}[0]{\scalebox{0.72}{\ensuremath{\mathbf{L}}}} 
\newcommand{\Time}[0]{\scalebox{0.72}{\ensuremath{\mathbf{T}}}} 
\newcommand{\Force}[0]{\scalebox{0.72}{\ensuremath{\mathbf{F}}}} 
\newcommand{\Energy}[0]{\scalebox{0.72}{\ensuremath{\mathbf{E}}}} 

\newcommand{\etaDilatational}{\ensuremath{\eta_{\emph{d}}}}
\newcommand{\etaSurface}{\eta_{\!\emph{s}}}

\newcommand{\limit}[2]{\ensuremath{\displaystyle\lim_{#1 \to #2}}}

\begin{document}

\title{Relaxation of a Simulated Lipid Bilayer Vesicle Compressed by an AFM}
\author{Ben M. Barlow, Martine Bertrand, and B\'ela Jo\'os}
\affiliation{Ottawa-Carleton Institute for Physics \\
University of Ottawa Campus\\ Ottawa, Ontario, Canada, K1N 6N5}
%\date{} % delete this line to display the current date

\graphicspath{{./images/}}

\begin{abstract}
\vspace*{0.2cm}
Using Coarse-Grained Molecular Dynamics simulations, we study the relaxation of bilayer vesicles, uniaxially compressed by an Atomic Force Microscope (AFM) cantilever. 
The relaxation time exhibits a strong force-dependence. 
Force-compression curves are very similar to recent experiments wherein giant unilamellar vesicles were compressed in a nearly identical manner. 
 \end{abstract}

\vspace*{0.2cm}
\date{\today}
\maketitle

% such that vectors are done in bold-face and unit vectors (typeset with \hat) are bold. 
\let\oldhat\hat
\renewcommand{\vec}[1]{\mathbf{#1}}
\renewcommand{\hat}[1]{\oldhat{\mathbf{#1}}}

%\linenumbers
\setstretch{2.0}
\section{Introduction} 
%▀▀▀▀▀▀▀▀▀▀▀▀▀▀▀▀▀▀▀▀▀▀▀▀▀▀▀▀▀▀▀▀▀▀▀▀▀▀▀▀▀▀▀▀▀▀▀▀▀▀▀▀▀▀▀▀▀▀▀▀▀▀▀▀▀▀▀▀▀▀▀▀▀▀▀▀▀▀▀▀▀▀▀▀▀▀▀▀▀▀▀▀▀▀▀▀▀▀▀▀▀▀▀▀▀▀▀▀▀▀
\noindent
Cells (the building blocks of life) are very complicated mechanical objects ---eukaryotic cells especially so. 
The plasma membrane, a lipid bilayer with many protein inclusions, separates the cell from the outside environment. 
As model physical systems, lipid bilayer vesicles (vesicles) have been an attractive starting point for theoretical work,  simulations and experiments. 
%They are used within cells and in the bloodstream. At its foundation, the cell membrane is a phospholipid bilayer. 
%Vesicles are integral to cell function, and knowledge of their mechanical properties and dynamics is important to understand living cells. 
Vesicles play an important role in cell function, e.g. storing and transporting substances throughout the cell. 
Their mechanical and dynamic properties are therefore of significance, not only for those functions, but also for the cell membrane whose basic structure is a lipid bilayer with a cytoskeleton and many inclusions. 
%%Vesicles serve as prototypes of simple cells and knowledge of their mechanical properties is important to understand living cells. 
%They serve as storage/transport containers, reaction vessels etc., and knowledge of their mechanical properties is important to understand a variety of biological processes. 

In this paper our focus is not on static properties, but on the dynamics of the stress relaxation. 
In particular we observe that the relaxation time depends on the magnitude of the applied stress, increasing sharply in the limit of low stress. 
Further, we show that this behaviour can be derived from the Helfrich and Servuss model\cite{helfrich_undulations_1984} for undulating elastic membranes. 
This derivation predicts a \emph{finite} maximum relaxation time, proportional to the membrane's surface area. 

\begin{figure}[h]
\includegraphics[width=0.98\textwidth]{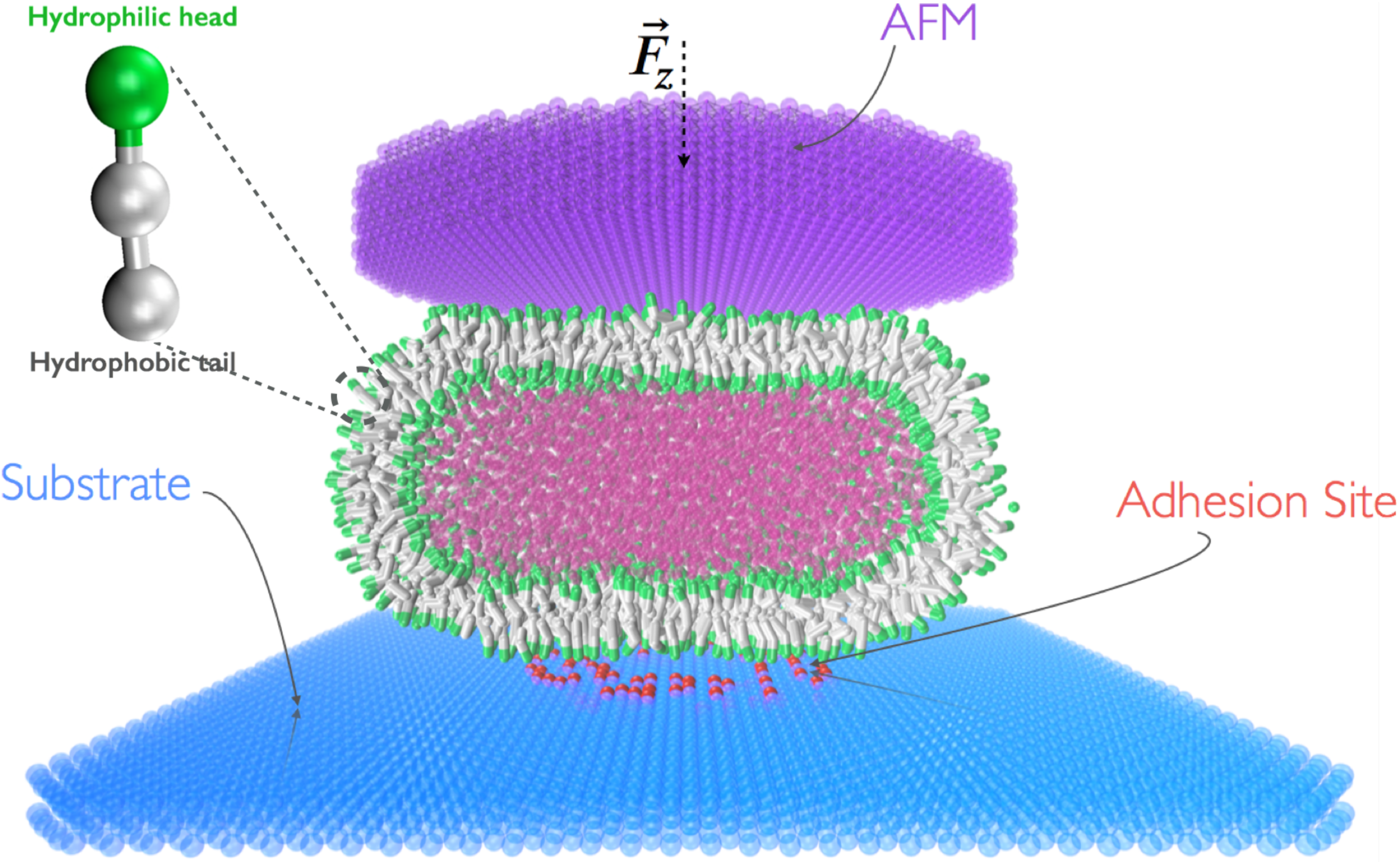} 
\caption[Simulated vesicle undergoing parallel plate compression.]{Simulated vesicle undergoing parallel plate compression. In addition to the ordinary substrate particles, a bullseye of randomly distributed `sticky' particles was placed at the centre of the substrate to ensure adhesion. Without this \emph{adhesion site}, the vesicle would slip out from underneath the AFM cantilever. Coarse-grained lipid shown at upper left.} 
\label{fig:introSqueezeVesicle} 
\end{figure}

To investigate the viscoelastic properties of vesicles, we ran computer simulations wherein a vesicle is squeezed between two plates (\autoref{fig:introSqueezeVesicle}). 
This procedure is relevant to experiments\cite{haase_resiliency_2013, schafer_mechanical_2013, schafer_mechanical_2015, guolla_force_2012, al-rekabi_cross_2013, hemsley_precisely_2011, silberberg_mitochondrial_2008} which use an Atomic Force Microscope (AFM) to poke and squeeze and stretch living cells and vesicles. 
An analogous experimental setup was used by Sch\"{a}fer et al.\cite{schafer_mechanical_2013} to investigate \emph{static} properties of  giant liposomes. 
But cells and vesicles are not deformed only in the lab. 
Inside our own bodies, every time the heart beats, every time we breathe, every time we flex a muscle of any kind  ---at every moment in cells all over the body, mechanical deformation of the membrane, cytoskeleton and cell contents is occurring. 
%\Martin{There are vesicles also in the body that deform as they are transported ...} 

In this paper, we show that the relaxation time of compressed vesicles increases sharply with decreasing force in the limit of small force (and low surface tension). 
In that limit the membrane exhibits significant undulations which are reduced by the squeezing of the vesicle. 
This entropic contribution to the relaxation time increases sharply as the force is decreased. Helfrich and Servuss\cite{helfrich_undulations_1984} (HS) have studied how membrane area expands with tension, and within their model we derive an expression for the relaxation time's force-dependence. 
The connection between our vesicle's relaxation time and the applied stress may help to explain the wide variability of relaxation (and recovery) times reported for cells. 
The maximum relaxation time scales as the membrane's surface area, so the force-dependence should be strong for cells and large vesicles as well. 
Scaled force-compression data is very similar to that reported for Giant Unilamellar Vesicles (GUVs) by Sch\"{a}fer et al.\cite{schafer_mechanical_2013}. 

\section{Model} 
%▀▀▀▀▀▀▀▀▀▀▀▀▀▀▀▀▀▀▀▀▀▀▀▀▀▀▀▀▀▀▀▀▀▀▀▀▀▀▀▀▀▀▀▀▀▀▀▀▀▀▀▀▀▀▀▀▀▀▀▀▀▀▀▀▀▀▀▀▀▀▀▀▀▀▀▀▀▀▀▀▀▀▀▀▀▀▀▀▀▀▀▀▀▀▀▀▀▀▀▀▀▀▀▀▀▀▀▀▀▀
\noindent
We use coarse-grained molecular dynamics simulations to reproduce the basic characteristics common to all real lipid bilayer membranes. 
The model (\autoref{fig:introSqueezeVesicle}) consists of approximately 140,000 particles in a simulation box with periodic boundary conditions. 
Our vesicle is the same as was used in \cite{bertrand_extrusion_2012} with reduced volume $\sim$ 1 (maximal volume without a pressure difference across the membrane), its membrane composed of coarse-grained lipids having one hydrophilic `head' particle and two hydrophobic `tail' particles. 
While relatively simple, these lipids are more than adequate for the present study. 
Our membrane exhibits thermal undulations, in-plane fluidity, intermonolayer friction, area compressibility, and bending rigidity, the basic features of fluid lipid bilayers. 
The model yields reasonable values for the area compressibility $\K$, and bending rigidity $\kappa$ (see \autoref{fig:AtriFit} and Section \ref{ssec:AlphaVsTension} as well as \cite{bertrand_extrusion_2012}). 
Despite the lipids' short chains, the membrane was not permeated by solvent, and lipid flips from one leaflet to the other were rare. 
We also note that there are advantages in using short lipids. 
There is the obvious reduction in simulation time, but the use of short lipids mitigates the disadvantages of small system size. 
Specifically, \emph{short lipids reduce the ratio of membrane thickness to vesicle diameter}. 
Said ratio decreases with vesicle size. 

The vesicles are constructed to attain a state where the internal and external fluid pressures are equal. 
The pressure difference has two contributions, potential and kinetic. 
The latter driven by temperature is significant and ensures that undulations persist in the bilayer up to lysis tension. 
At 3000 lipids, the membrane area is $\sim 10^{3}$  times the area per lipid, large enough to achieve the macroscopic properties described by continuum models. 
\autoref{fig:introSqueezeVesicle} omits the outer fluid particles surrounding our small unilamellar vesicle. 
The explicit solvent filling and surrounding the vesicle is a Lennard-Jones fluid, at an initial density of $0.8 \text{ particles}/\Length^{3}$. 
($\Length$ is the unit of length, introduced in \autoref{ss:Units}.) 
The vesicle is sandwiched between a substrate and an AFM cantilever ---both consisting of fluid-like particles, constrained to remain in an fcc lattice. 
On the scale of our simulations, we treat a rounded AFM tip as approximately flat. 
For giant vesicles, this corresponds to a tipless AFM cantilever. 

In molecular dynamics, pair potentials are defined which determine the force exerted by each particle on its neighbours and vice-versa. 
To include thermal motion, there is an additional random force applied to each particle (generated by the simulation's `thermostat'). 
With the force on each particle determined by the thermostat and pair potentials, the time evolution of the system is governed by Newton's laws of motion. 
Our system is in the $NVT$ ensemble, simulated using HOOMD-blue\cite{anderson_general_2008, HOOMDwebsite} with a DPD thermostat\cite{phillips_pseudo-random_2011}.
The DPD thermostat uses pairwise interactions to rescale particle velocities, which means that not only temperature is kept constant, but momentum is conserved ---necessary for dynamic processes like the relaxation simulated here. 
The initial state was prepared using both the ESPResSo\cite{arnold_espresso_2013, ESPResSowebsite} and HOOMD-blue simulation packages. 
The Python packages matplotlib\cite{hunter_matplotlib:_2007, matplotlibwebsite}, MDAnalysis\cite{michaud-agrawal_mdanalysis:_2011, MDAnalysiswebsite} and Numpy \& Scipy\cite{van_der_walt_numpy_2011, NumpyScipywebsite} were used to plot and analyze our data. 
\subsection{Potentials} 
%▔▔▔▔▔▔▔▔▔▔▔▔▔▔▔▔▔▔▔▔▔▔▔▔▔▔▔▔▔▔▔▔▔▔▔▔▔▔▔▔▔▔▔▔▔▔▔▔▔▔▔▔▔▔▔▔▔▔▔▔▔▔▔▔▔▔▔▔▔▔▔▔▔▔▔▔▔▔▔
\noindent
The two key interaction potentials in our simulation are the Lennard-Jones potential
\begin{linenomath*}\begin{equation} \VLJ = 4\Energy\left[\left(\frac{\Length}{r}\right)^{12}  -  s\left(\frac{\Length}{r}\right)^{6}\right]\text{,}   \label{eqn:LJmodified}\end{equation}\end{linenomath*} 
and the soft-sphere  potential 
\begin{linenomath*}\begin{equation} \VSS = ar^{-9} \text{.} \label{eqn:SoftSpherePotential}\end{equation}\end{linenomath*} 
$\Energy$ is the unit of energy, introduced in \autoref{ss:Units}. 
$r$ is the particle separation. 
$s$ is a parameter allowing the strength of the attractive portion of $\VLJ$ to be tuned (default is $s = 1$). 
$a$ tunes the strength of the soft-sphere (hydrophobic) potential.

$\VLJ$ with $s = 1$ governs all non-bonded interactions between same-type particles, and between most particles of different types. 
The key exception is the hydrophobic tail-fluid interaction, which is governed by $\VSS$. 

Bonds between monomers in the coarse-grained lipids are governed by a harmonic potential 
\begin{linenomath*}\begin{equation} 
\label{harmonic bond} V_{\emph{harm}} = \frac{k}{2}(r - r_{0} )^{2} \text{,}  
\end{equation}\end{linenomath*} 
with $k = 5000 \Energy/\Length^{2}$. 
Bonds among particles making up the AFM probe, as well as bonds between the substrate particles and their anchor points are implemented using $V_{\emph{harm}}$ with $k=3000 \Energy/\Length^{2}$.

These are the same potentials as were used in \cite{goetz_computer_1998, bertrand_extrusion_2012}, plus two additional potentials. % ---one to constrain the motion of the AFM Tip, the other to adhere the vesicle to the substrate. 
First of the two is a cylindrical harmonic potential. 
This potential is used to keep the AFM centred and level during compression, by constraining its constituent particles to vertical motion. 
(The entire crystal is effectively \emph{riding on rails}.) 
Second is $\VLJ$ but with $s>1$ ---the strength of the attractive term increased using the $s$-parameter. 
This latter pair potential was used for the interaction of the `adhesion site' (in the centre of the substrate, coloured red in \autoref{fig:introSqueezeVesicle}) with the lipid heads. 
The enhanced attraction causes the lipid heads to stick to the adhesion site, keeping the vesicle centred under the AFM.

\subsection{Units}\label{ss:Units} 
%▔▔▔▔▔▔▔▔▔▔▔▔▔▔▔▔▔▔▔▔▔▔▔▔▔▔▔▔▔▔▔▔▔▔▔▔▔▔▔▔▔▔▔▔▔▔▔▔▔▔▔▔▔▔▔▔▔▔▔▔▔▔▔▔▔▔▔▔▔▔▔▔▔▔▔▔▔▔▔

We denote our simulations' dimensionless units $\Length =$ length, $\Mass =$ mass, $\Time =$ time, $\Energy =$ energy, and $\Force =$ force. 
Figure axes are in dimensionless units when no S.I. units are specified. 
The conversion to dimensionful units (detailed in Appendix \ref{app:UnitsConversion}) yields

%\begin{linenomath*}\begin{equation} \begin{cases} 
%\Length \approx  0.5 \nano\metre  \\
%\Mass \approx  6\times 10^{-26}\kilo\gram & \approx  \left(\substack{\text{mass of 2 water molecules,} \\ \text{or 3 carbon atoms}}\right)  \\
%\Time \approx  2\pico\second \\
%\Energy \approx  4\times 10^{-21}\Joule & \substack{\text{thermal energy per particle at} \\ \text{room temperature ($T = 25^{\circ}\textrm{C}$)}} \\
%\Force \approx  10 \pico\Newton\text{.} 
%\end{cases}    \label{eqn:units}\end{equation}\end{linenomath*} 

\begin{linenomath*}\begin{equation} \begin{cases} 
\Length \approx  0.6 \nano\metre  \\
\Mass \approx  8.5\times 10^{-26}\kilo\gram & \approx  \left(\substack{\text{mass of 3 water molecules,} \\ \text{or 4 carbon atoms}}\right)  \\
\Time \approx  3\pico\second \\
\Energy \approx  4\times 10^{-21}\Joule & \substack{\text{thermal energy per particle at} \\ \text{room temperature ($T = 25^{\circ}\textrm{C}$)}} \\
\Force \approx  6.6\pico\Newton\text{.} 
\end{cases}    \label{eqn:units}\end{equation}\end{linenomath*} 

These unit conversions are meant only as a rough guide to help scale the simulation in the context of lipid bilayer vesicles. 
If, instead of a vesicle, we were mapping our simulation to some other physical system, then different unit conversions would be invoked. 
(The validity of a given computer simulation might extend beyond the original system being studied.)

\section{Results} 
%▀▀▀▀▀▀▀▀▀▀▀▀▀▀▀▀▀▀▀▀▀▀▀▀▀▀▀▀▀▀▀▀▀▀▀▀▀▀▀▀▀▀▀▀▀▀▀▀▀▀▀▀▀▀▀▀▀▀▀▀▀▀▀▀▀▀▀▀▀▀▀▀▀▀▀▀▀▀▀▀▀▀▀▀▀▀▀▀▀▀▀▀▀▀▀▀▀▀▀▀▀▀▀▀▀▀▀▀▀▀
\noindent
\subsection{Relaxation time versus force} 
\begin{figure}[h]
\includegraphics[width=0.98\textwidth]{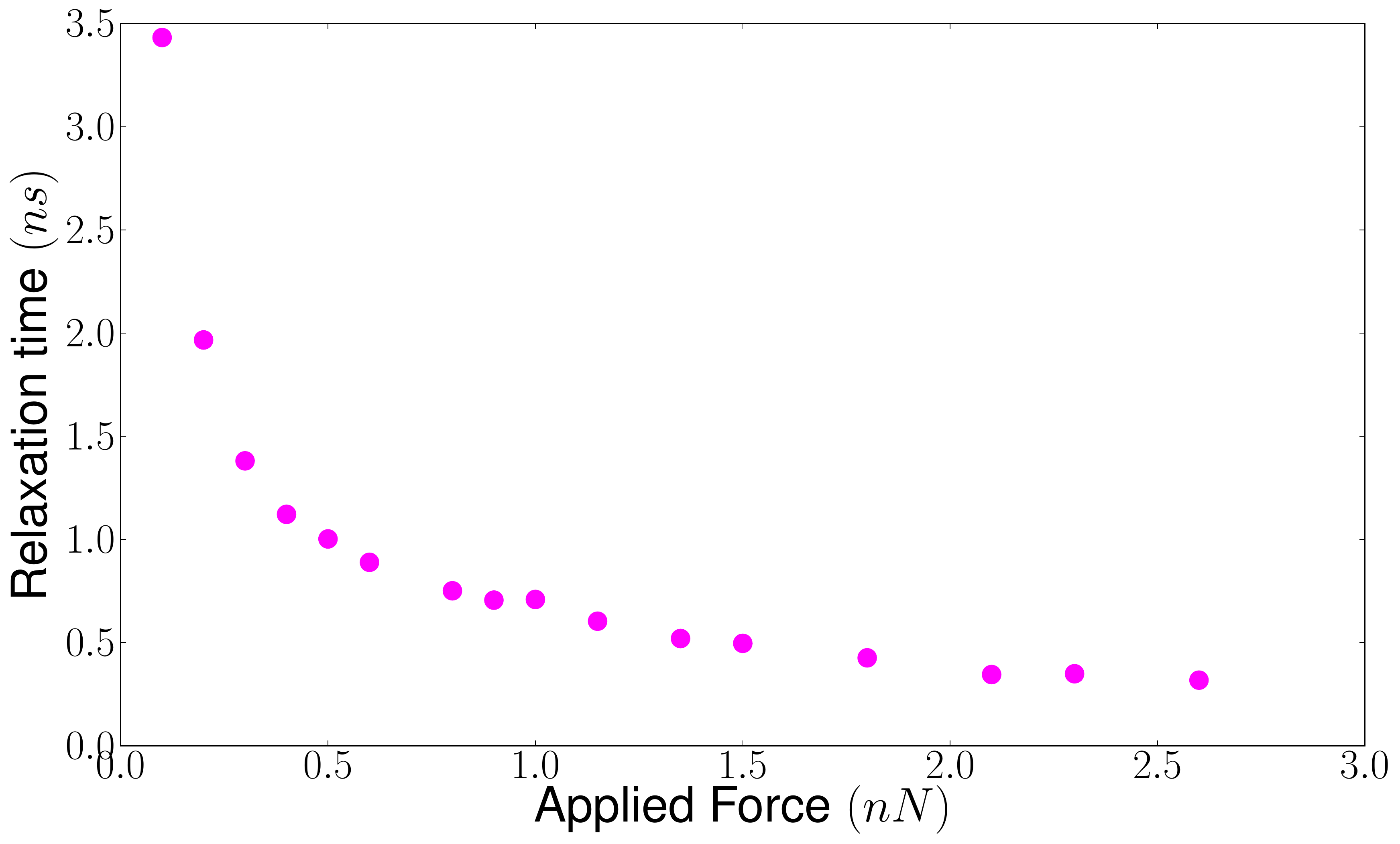} 
\caption[Relaxation time versus force]{Relaxation time plotted as a function of the squeezing force. The relaxation time shows a steep increase at low force. (Units are approximate, see Appendix \ref{app:UnitsConversion}.)} 
\label{fig:tauVsForceNoFit} 
\end{figure}

%▔▔▔▔▔▔▔▔▔▔▔▔▔▔▔▔▔▔▔▔▔▔▔▔▔▔▔▔▔▔▔▔▔▔▔▔▔▔▔▔▔▔▔▔▔▔▔▔▔▔▔▔▔▔▔▔▔▔▔▔▔▔▔▔▔▔▔▔▔▔▔▔▔▔▔▔▔▔▔
\noindent
When we squeeze the vesicle, it relaxes to a new steady state with a characteristic time constant $\tau$, which we call the relaxation time. 
We calculate this quantity by following the time evolution of the area expansion. 
A key result shown in \autoref{fig:tauVsForceNoFit} is that the \emph{relaxation time depends strongly on the applied stress}, showing a sharp increase at low force.

We explain this result in \autoref{derHelfrich} using the HS model\cite{helfrich_undulations_1984}. 
The sharp rise in the vesicle's relaxation time at low force arises from the effect of entropic undulations on the area expansion. 

\begin{figure}[h]
\includegraphics[width=0.98\textwidth]{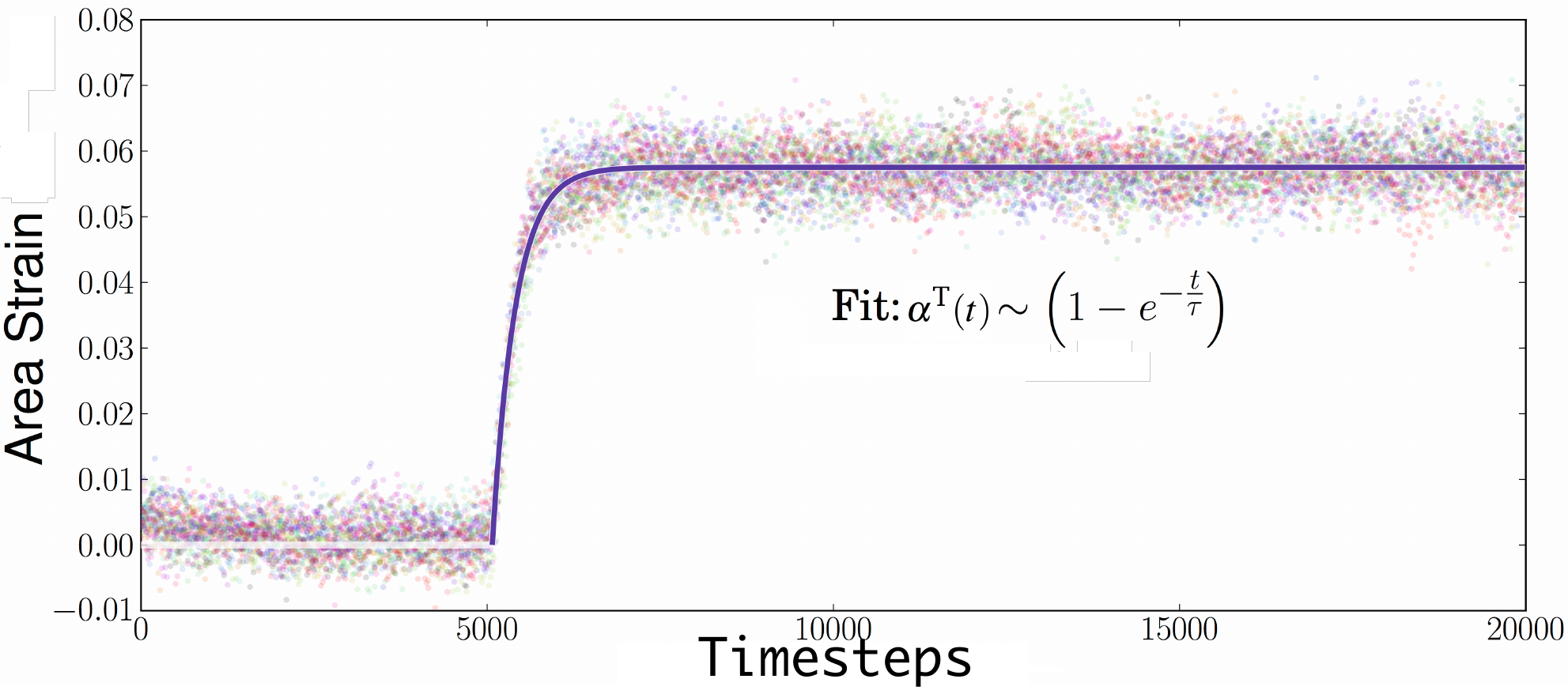} 
\caption[Ensemble fit to creep response of bilayer]{Ensemble fit to creep response of the bilayer's triangulated area at $F_{\!z} =-100\Force$. For each value of the applied force, data from multiple simulations are fit as one timeseries. This helps to reduce the uncertainty on the relaxation time, by reducing the influence of noise from any particular simulation on the fit.}\label{fig:alphaVsTime} 
\end{figure}

\begin{figure}[h]
\includegraphics[width=0.98\textwidth]{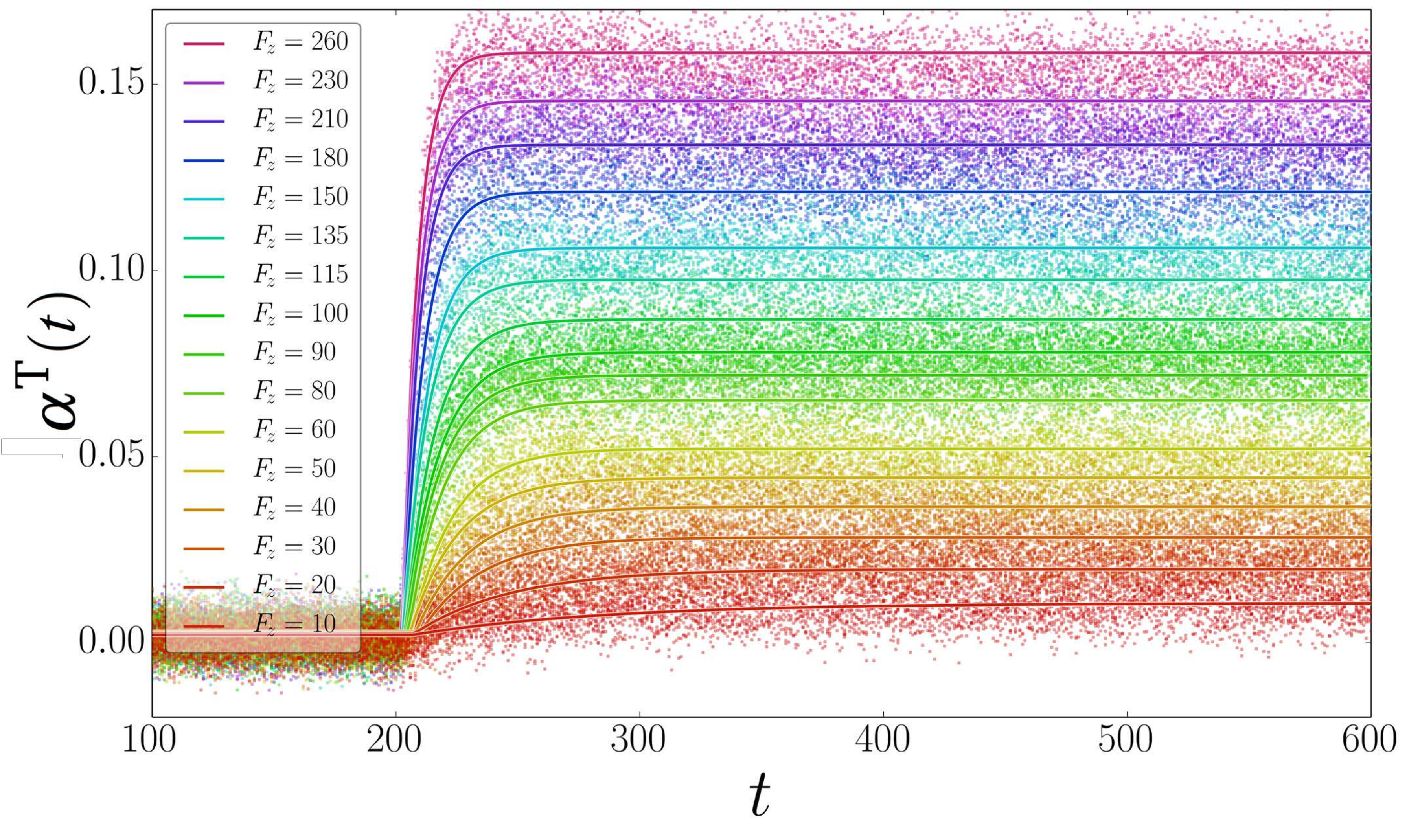} 
\caption[Triangulated area expansion timeseries of vesicle at different forces]{Triangulated area expansion timeseries of vesicle at different forces (colour online). Squeezing force is indicated by colour, with red$ = 10\Force$ and violet$ = 260\Force$. Solid lines indicate fits to triangulated area $\alpha (t)$ from which relaxation times are obtained.} 
\label{fig:alphaVsTimeEmptyVesicle} 
\end{figure}

The time evolution of the area strain $\alpha$ after we activate the squeezing force is described as an exponential saturation
\begin{linenomath*}\begin{equation}  \alpha (t)  = \alpha_{\infty}\left(1 - e^{-{t}/{\tau}}\right) \text{,}\label{eqn:SLSmodel}\end{equation}\end{linenomath*} 
as illustrated in Figures \ref{fig:alphaVsTime} and \ref{fig:alphaVsTimeEmptyVesicle}. 
This type of viscoelastic creep response corresponds to the `Kelvin-Voigt' model, or the more general `Standard Linear Solid' (SLS) model. 
In this model the relaxation time is $\tau \sim {\eta/K}$, where $\eta$ is a viscosity and $K$ is an elastic modulus.

The \emph{triangulated} area strain $\alphaTriangulated$ of the vesicle (Figures \ref{fig:alphaVsTime} and \ref{fig:alphaVsTimeEmptyVesicle}) was used to obtain the relaxation times. 
This relative area change is calculated with a script used previously in\cite{bertrand_extrusion_2012, bertrand_deformed_2012}, which implements Nina Amenta's `crust' algorithm \cite{amenta_new_1998} to triangulate the inner and outer leaflets of the vesicle (see Appendix \ref{ssec:TriangulatedSurface}). 
In all further analysis, the apparent area of the vesicle (or ``projected area'') was used, as it is more amenable to modeling. 
There is therefore the assumption that the relaxation time does not depend on the specific way the area is calculated.     

The full fitting function used in Figures \ref{fig:alphaVsTime} and \ref{fig:alphaVsTimeEmptyVesicle} is
\begin{linenomath*}\begin{equation} 
\alphaTriangulated(t) =  
\begin{cases}  \alphaTriangulated_{0}  &\text{for $t < t_{0}$} \\ 
\alphaTriangulated_{0} + \alphaTriangulated_{\infty}\left( 1 - e^{-(t - t_{0})/\tau} \right)  &\text{for $t \geq t_{0}$.} 
 \end{cases} 
 \label{eqn:fittingFunction} 
 \end{equation}\end{linenomath*} 
Combining the creep response at $t \geq t_{0}$ with a flatline at $t < t_{0}$ ---the initial time $t_{0}$ being a free parameter---  gives a more robust fit. 
%This combination of a creep response at $t \geq t_{0}$ with a flatline at $t<t_{0}$ gives a more robust fit. 

 Even at relatively high forces ($>100\Force$), fluctuations in the vesicle's surface area are fairly large ---on the same order of magnitude as the mean area expansion. 
For this reason, when fitting for the relaxation time at a given force, data from multiple equivalent simulations are superposed and then fit as a single timeseries (see \autoref{fig:alphaVsTime}). 
That is, the relaxation time is fit to an ensemble of simulations. 
This way, the influence of random fluctuations from any particular timeseries is reduced.

\subsection{Projected area expansion versus force} 
%▔▔▔▔▔▔▔▔▔▔▔▔▔▔▔▔▔▔▔▔▔▔▔▔▔▔▔▔▔▔▔▔▔▔▔▔▔▔▔▔▔▔▔▔▔▔▔▔▔▔▔▔▔▔▔▔▔▔▔▔▔▔▔▔▔▔▔▔▔▔▔▔▔▔▔▔▔▔▔
\noindent

Due to thermal undulations, the surface area of a vesicle as measured in the lab will be less than the true surface area of its membrane. 
What one actually measures is the surface area of an \emph{apparent surface} ---the surface one gets by smoothing over the rapid fluctuations in membrane shape (see Appendix \ref{app:ApparentSurface}). 
That is why the distinction is made between `apparent' or `projected' versus true surface area of the membrane. 

\begin{figure}[h]
\includegraphics[width=0.98\textwidth]{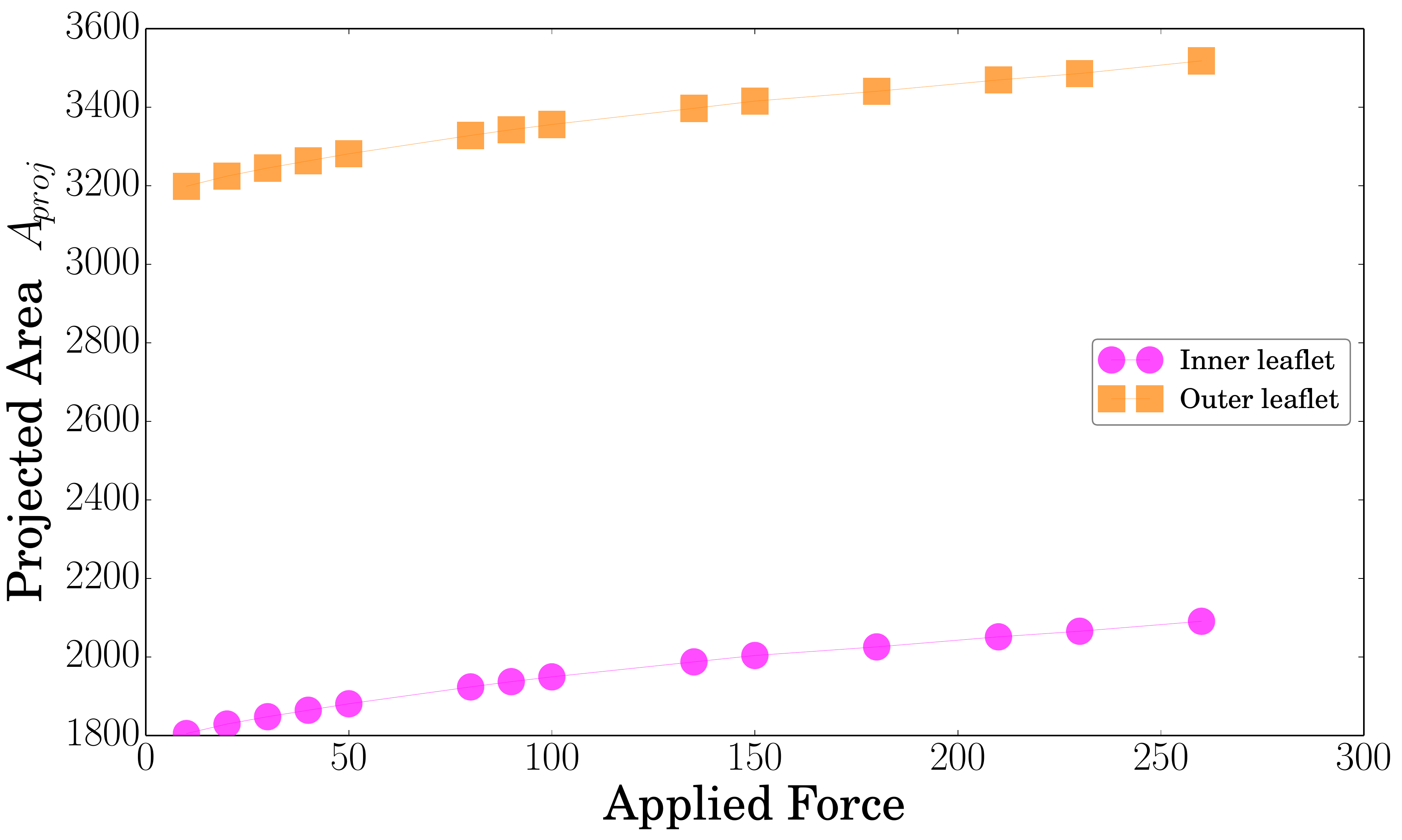} 
\caption[Projected Area versus force]{Projected Area (area of apparent surface, see Appendix \ref{app:ApparentSurface}) of the bilayer versus force. Each leaflet is plotted separately.} 
\label{fig:projectedAreaVsForce} 
\end{figure}

\autoref{fig:projectedAreaVsForce} shows our simulated vesicle's projected surface area versus force. 
Both leaflets are shown. 

The increase is logarithmic at low force and linear at high force.

\subsection{The Helfrich-Servuss (HS) model} 
%▔▔▔▔▔▔▔▔▔▔▔▔▔▔▔▔▔▔▔▔▔▔▔▔▔▔▔▔▔▔▔▔▔▔▔▔▔▔▔▔▔▔▔▔▔▔▔▔▔▔▔▔▔▔▔▔▔▔▔▔▔▔▔▔▔▔▔▔▔▔▔▔▔▔▔▔▔▔▔
%\noindent
To establish a physical basis for the force-dependence of the relaxation time $\tau$ (\autoref{fig:tauVsForceNoFit}), 
we begin by introducing the HS model\cite{helfrich_undulations_1984}, which is used in the next section to derive an expression for the relaxation time as a function of tension. 

Membranes behave as entropic springs. 
Thermal agitation excites undulations in the vesicle membrane. 
If $A$ is the zero temperature area of the membrane, the undulations will reduce the apparent or projected area by $\Delta A$ (negative at zero tension $\gamma$). 
When the membrane tension $\gamma$ is increased, these undulations are reduced ($\Delta A$ approaches zero).  
This flattening of undulations by surface tension reduces the number of microstates (shapes) available to the vesicle, decreasing its entropy ---just as pulling the ends of an entropic spring reduces the number of states available to it. 
If we increase the surface tension beyond the point at which undulations are largely suppressed,  direct stretching of the membrane dominates. 
This true stretching is called `direct area expansion' whereas flattening undulations increases the apparent surface area of the membrane without actually stretching it. 
The \emph{observed} area (a.k.a. `apparent' or `projected' area) expansion results from a combination of these two effects.

In 1984 Helfrich \emph{\&} Servuss \cite{helfrich_undulations_1984} derived an expression 
relating the relative change in a membrane's projected area $\Delta A$ to its surface tension $\gamma$:
\begin{linenomath*}\begin{equation} \alpha(\gamma) \equiv \left(\frac{\Delta A}{{A}}\right)_{\gamma >0}  = \underbrace{\frac{\kBoltzmann T}{8\pi \kappa }\ln{\left(\frac{\frac{\zeta }{A} + \frac{\gamma }{appa }}{\frac{\zeta }{a} + \frac{\gamma }{\kappa }}\right)}}_{\emph{entropic}} + \underbrace{\frac{\gamma }{\K}}_{\emph{direct}} \text{,} \label{eqn:alphaHelfrich}\end{equation}\end{linenomath*} 
where $\K$, $\kappa$, $A$ and $a$ are the membrane's area compressibility modulus, bending rigidity, unstressed area and area per lipid, respectively. 
$\kBoltzmann $ is Boltzmann's constant, $T$ is the temperature,  and $\zeta$ is a parameter which depends on membrane shape. (E.g. $\zeta = \pi^{2}$ for a planar membrane, and for a sphere $\zeta = 24\pi$.)
A mnemonic for \autoref{eqn:alphaHelfrich} is 
\begin{linenomath*}\begin{equation}  \alpha(\gamma) = \alphaEntropic(\gamma) + \alphaDirect(\gamma)  \text{;} \label{eqn:alphaHelfrichMnemonic}\end{equation}\end{linenomath*}  
where $\alphaEntropic(\gamma)$, the first term in \autoref{eqn:alphaHelfrich}, is negative (tending to zero as $\gamma \rightarrow \infty$) since it measures the portion of membrane area $A$ absorbed by undulations. 

Vesicle size matters: larger membranes have more of their surface area hidden in undulations. 
In other words, $\alphaEntropic(\gamma)$ is more negative for larger $A$ (see \autoref{eqn:alphaHelfrich}). 
Since wavelengths present in the bilayer can't exceed the vesicle circumference, the spectrum of undulations is constrained by vesicle size. 
In fact, as $\gamma \rightarrow 0$ the undulations' mean square amplitude (which is dominated by the longest wavelengths present) scales as the membrane area.\cite{helfrich_undulations_1984}

Evans and Rawicz\cite{evans_entropy-driven_1990} studied the area expansion of vesicles subject to tensions $10^{-7}\leq \gamma \leq 10^{-3} \Newton/\metre$, and observed a logarithmic dependence at low tension followed by a linear dependence at larger tensions ---consistent with the HS model. 
Further empirical support for the HS model was provided by Dimova et al.\cite{dimova_vesicles_2009}. 
In this case, GUVs were deformed using electric fields, and their area expansion plotted against the resulting membrane tension. 

More recent experiments (see Figure 2 of  Mell et al.\cite{mell_fluctuation_2015}) have shown that the undulation spectrum $P = P\HS(\ell, \gamma)$ (\autoref{eqn:HSspectrum}) used in deriving the HS model\cite{helfrich_undulations_1984} departs from experimental spectra at high wavenumber $\ell$.
In Appendix \ref{app:RevisedHSmodel} we use the spectrum $P\Mell(\ell, \gamma)$ (\autoref{eqn:PMell}) to derive a `revised HS model':
\begin{linenomath*}
\begin{equation}  \alpha(\gamma) \approx \underbrace{\frac{\kBoltzmann T}{8\pi \kappa }\ln{\left(\frac{\frac{\zeta }{A} + \frac{\gamma }{\kappa }}{\frac{\zeta }{a} + \frac{\gamma }{\kappa }}\right)} + \frac{3\kBoltzmann T}{\pi h^{2} \K }\ln{\left(\frac{1 + \frac{hR\zeta }{2A}}{1 + \frac{hR\zeta }{2a}}\right)}}_{\emph{entropic}} + \underbrace{\frac{\gamma }{\K}}_{\emph{direct}} \text{.}\label{eqn:RevisedHelfrich}\end{equation}
\end{linenomath*} 
The above correction alters $\alphaEntropic$ in \autoref{eqn:alphaHelfrich}, shifting it by a term which is independent of the surface tension. 
Being independent of the tension, this correction does not alter our model for the relaxation time, as we will see below.

\subsection{Derivation of relaxation time} \label{derHelfrich} 
%▔▔▔▔▔▔▔▔▔▔▔▔▔▔▔▔▔▔▔▔▔▔▔▔▔▔▔▔▔▔▔▔▔▔▔▔▔▔▔▔▔▔▔▔▔▔▔▔▔▔▔▔▔▔▔▔▔▔▔▔▔▔▔▔▔▔▔▔▔▔▔▔▔▔▔▔▔▔▔
We now proceed to derive the relaxation time using the HS model as the starting point. 
From linear viscoelasticity theory, we have \begin{linenomath*}\begin{equation} \tau  \sim   \frac{\eta}{K} \text{.}   \label{eqn:tauViscosityOverModulus}\end{equation}\end{linenomath*} 
So the viscosity $\eta$ and elastic modulus $K$ need to be specified. 
The most physically appropriate viscosity is called the \emph{dilatational-surface viscosity}\cite{dimova_giant_2007, riske_electro-deformation_2005} $\etaDilatational $ ---the viscosity associated with stretching the membrane, which we assume to be $\approx $ constant so that \begin{linenomath*}\begin{equation} \tau  \propto   \frac{1}{K} \text{.}  \label{eqn:tauPropToOneOverK}\end{equation}\end{linenomath*}

\noindent
To obtain $K$ we return to the heart of elasticity theory. 
Hooke's law suggests a more general definition for $K$:
For small $\Delta (\cdot)$ we know that
\begin{linenomath*}\begin{equation}  \Delta (\emph{strain}) = \frac{1}{K}\Delta(\emph{stress}) \approx  \left(\frac{\partial (\emph{strain})}{\partial (\emph{stress}) }\right)\Delta (\emph{stress})\text{.} \label{eqn:strain=StressOverK}\end{equation}\end{linenomath*} 
In the case of a stretching membrane $\emph{strain} = \alpha $ (relative increase in the apparent area) and $\emph{stress} = \gamma $ (the surface tension), so that
\begin{linenomath*}\begin{equation}  \frac{1}{K}  \equiv \frac{\partial (\emph{strain})}{\partial (\emph{stress}) }  = \frac{\partial \alpha }{\partial \gamma }   \label{eqn:Kmembrane}\end{equation}\end{linenomath*} 
defines the effective modulus $K$ of the bilayer (in the vicinity of a specific value of $\gamma $).

With $\alpha(\gamma) $ specified by \autoref{eqn:RevisedHelfrich} (which turns out to be equivalent to \autoref{eqn:alphaHelfrich} in our case, since the second entropic term does not depend on $\gamma$), 
%\footnote{The correction to the HS model (middle term in \autoref{eqn:RevisedHelfrich}) that results from $P\Mell$\cite{mell_fluctuation_2015} does not affect this derivation ---see Appendix \ref{app:RevisedHSmodel}.}
\autoref{eqn:Kmembrane} yields
\begin{linenomath*}\begin{equation}\frac{1}{K}  = \frac{1}{\K} + \underbrace{\frac{\kBoltzmann T}{8\pi\kappa}}_{\text{``$M$''}}\left\{\frac{1}{\frac{\zeta \kappa }{A} + \gamma } - \frac{1}{\frac{\zeta \kappa }{a} + \gamma }\right\}\text{.}   \label{eqn:oneOverKFullForm}\end{equation}\end{linenomath*} 
Note that \autoref{eqn:oneOverKFullForm} includes temperature, bending modulus, and area compressibility.

Returning to \autoref{eqn:tauViscosityOverModulus}, which relates  relaxation time, viscosity and elasticity, \autoref{eqn:oneOverKFullForm} predicts (via the HS model) a relaxation time 
\begin{linenomath*}\begin{equation}  \tau (\gamma ) \sim  \frac{\eta }{K} \approx  \eta \left(\frac{1}{\K}  + \frac{M}{\frac{\zeta \kappa }{A} + \gamma } + \frac{M}{\frac{\zeta \kappa }{a} + \gamma }\right) \text{.} \label{eqn:tauGammaHelfrich}\end{equation}\end{linenomath*} 

Since $a \ll A$ and our simulations occur in the regime $\gamma \ll \frac{\zeta\kappa}{a}$, the term $\frac{1}{\frac{\zeta\kappa}{a} + \gamma} \approx  \frac{a}{\zeta\kappa}$ and can be dropped from \autoref{eqn:oneOverKFullForm}. 
\autoref{eqn:tauGammaHelfrich} then simplifies to 
\begin{linenomath*}\begin{equation}  \tau (\gamma ) \sim  \frac{\eta }{K} \approx  \eta \left(\frac{1}{\K}  + \frac{M}{\frac{\zeta \kappa }{A} + \gamma } \right) \text{.} \label{eqn:tauGammaHelfrich2}\end{equation}\end{linenomath*} 

%###################################################################################
The relaxation time approaches a \emph{finite} limit as the tension vanishes, and at high tension it decreases asymptotically toward $(\eta/\K)$ :
\begin{linenomath*}\begin{equation} 
\begin{cases} 
\tau (\gamma ) \approx  \eta \left(\frac{1}{\K}  + \frac{MA}{\zeta \kappa }\right)  &\text{for vanishing tension, and}  \\\\
\tau (\gamma ) \approx   \eta \left( \frac{1}{\K} + \frac{M}{\gamma } \right)            &\text{for larger tensions, i.e. $\gamma  \gg  \frac{\zeta \kappa }{A}$.} 
\end{cases} 
\label{eqn:tauGammaBothLimits}\end{equation}\end{linenomath*} 
The low-tension limit of $\tau$ increases as the surface area of the membrane, predicting longer relaxation times for larger vesicles and cells at low tension. 
The high-tension limit agrees with the observation by Dimova et al.\cite{dimova_giant_2007, riske_electro-deformation_2005} that for giant vesicles near lysis tension $\tau \sim  \frac{\eta }{\gamma } $. 
Their result was justified through dimensional analysis. 
%###################################################################################

A phenomenological form consistent with both the low and high tension limits (\autoref{eqn:tauGammaBothLimits}) is 
\begin{linenomath*}\begin{equation} \tau   \approx   C_{1} + \frac{C_{2}}{C_{3} + \gamma }    \text{,}  \label{eqn:tauGammaHelfrichFit}\end{equation}\end{linenomath*} 
where $C_{1}$ is the high-tension asymptotic limit and $C_{1} + \frac{C_{2}}{C_{3}} $ is the finite limit as $\gamma \rightarrow 0$. 
At low tension, the vesicle shape remains nearly spherical. 
At high tension, the vesicle shape is again approximately constant, this time resembling a wheel of cheese. 
So at both limits $\zeta \approx \text{constant}$, and \autoref{eqn:tauGammaHelfrichFit} (derived from the HS model) is valid. 
Going a step further, in \autoref{ssec:TauVsTension} we fit the entire $\tau(\gamma)$ curve with this function, which succeeds as a phenomenological model and yields an estimate of $ (\eta/\K)$.

\subsection{Tension versus force} \label{ssec:TensionVsForce} 
%▔▔▔▔▔▔▔▔▔▔▔▔▔▔▔▔▔▔▔▔▔▔▔▔▔▔▔▔▔▔▔▔▔▔▔▔▔▔▔▔▔▔▔▔▔▔▔▔▔▔▔▔▔▔▔▔▔▔▔▔▔▔▔▔▔▔▔▔▔▔▔▔▔▔▔▔▔▔▔
\noindent
In the foregoing analysis we arrived at a model for the vesicle's relaxation time $\tau$ \emph{as a function of the surface tension $\gamma$} (Equations \ref{eqn:tauGammaHelfrich2}--\ref{eqn:tauGammaHelfrichFit} above). 
The goal now is to apply that model to the simulated vesicle. 
However, our simulation data gave the relaxation time \emph{as a function of the squeezing force  $\F$} (\autoref{fig:tauVsForceNoFit}), not of the tension. 
(The tension in the membrane is not an explicit parameter of our MD simulations, but rather is an \emph{effect} of the squeezing force $\F$.) 
We therefore \emph{need to know how $\gamma$ varies as a function of $\F$}. 

The surface tension $\gamma(\F)$ was calculated from the differential work $\d W$ done in deforming the vesicle. 
At each value of the squeezing force the vesicle was allowed to equilibrate, then the projected area, pressure and volume were measured (e.g. \autoref{fig:projectedAreaVsForce}). 
To approximate the surface tension at equilibrium as a function of the force, these measurements were used to obtain the tension from a relationship between equilibrium quantities, so the approximation of quasi-static deformation is applicable (Equations \ref{eqn:dF=dW}, \ref{eqn:tensionFreeEnergy}). 
%(For a detailed discussion, see \cite{barlow_poking_2015}.)
Since forms of deformation other than area expansion also contribute to $\d {W}$, the contribution due to $\gamma $ had to be extracted from the total work.

Because our system is $NVT$, the differential mechanical work $\d W$ done by the AFM (while squeezing the vesicle) is equal to the change in the system's free energy $\d\mathcal{F}$:
\begin{linenomath*}\begin{equation}  T = \emph{const.} \implies \d \mathcal{F} = \d W \text{.} \label{eqn:dF=dW}\end{equation}\end{linenomath*} 

%\noindent As such, the surface tension $\gamma $ can be defined in terms of the differential free energy
\noindent This is useful, since the surface tension $\gamma $ can be defined in terms of the differential free energy
\begin{linenomath*}\begin{equation} \d \mathcal{F} = \gamma \d A - \displaystyle\sum\limits_{j}P_{j}\d V_{j} \label{eqn:tensionFreeEnergy}\end{equation}\end{linenomath*} 
of the system (i.e. vesicle and solvent). 
%That is, the surface tension measures the change in free energy needed to increase the membrane's surface area by $\d A$. 
The sum over $j$ reads \begin{linenomath*}\begin{equation}  \displaystyle\sum\limits_{j}\Big(\Big)_{j} = \Big(\Big)_{\emph{inner}\cdot \emph{fluid}} + \Big(\Big)_{\emph{membrane}}+ \Big(\Big)_{\emph{outer}\cdot \emph{fluid}} \text{.}    \label{eqn:sumOverPjdVj}\end{equation}\end{linenomath*}

The $\gamma \d A$ term is the work done increasing the area of the membrane, and the sum over $P_{j}\d V_{j}$ accounts for other work which may be done compressing/expanding the volume of the inner/outer fluid and of the membrane. 
Combining Equations \ref{eqn:dF=dW} and \ref{eqn:tensionFreeEnergy} and dividing by $\d A$ gives
\begin{linenomath*}\begin{equation} \gamma  =  \frac{dW}{dA} + \displaystyle\sum\limits_{j}P_{j}\frac{\d V_{j}}{\d A } \text{.} \label{eqn:tensionWorkDefinition}\end{equation}\end{linenomath*} 
Everything on the right hand side of \autoref{eqn:tensionWorkDefinition} is a function of $\F$ ---the squeezing force. 
The $P_{j}(\F)$, the $\d V_{j}(\F)$, $\d A(\F)$ and $\d z(\F)$ are obtained by curve-fitting (then numerically differentiating) the pressures, volumes, area, and AFM cantilever height (respectively) as functions of $\F$. 
(Various regions' volumes and the membrane area are obtained by curve-fitting the vesicle's inner and outer surfaces as explained in Appendix \ref{app:ApparentSurface}.) 
Knowing $\d z(\F)$ and $\d A(\F)$ also takes care of
\begin{linenomath*}\begin{equation}   \frac{dW}{dA} =  \F\left(\frac{dz}{dA}\right) \text{,}  \label{eqn:dWdA}\end{equation}\end{linenomath*} 
completing \autoref{eqn:tensionWorkDefinition}.

\noindent

\subsection{Relaxation time versus tension} \label{ssec:TauVsTension} 
\begin{figure}[h]
\includegraphics[width=0.98\textwidth]{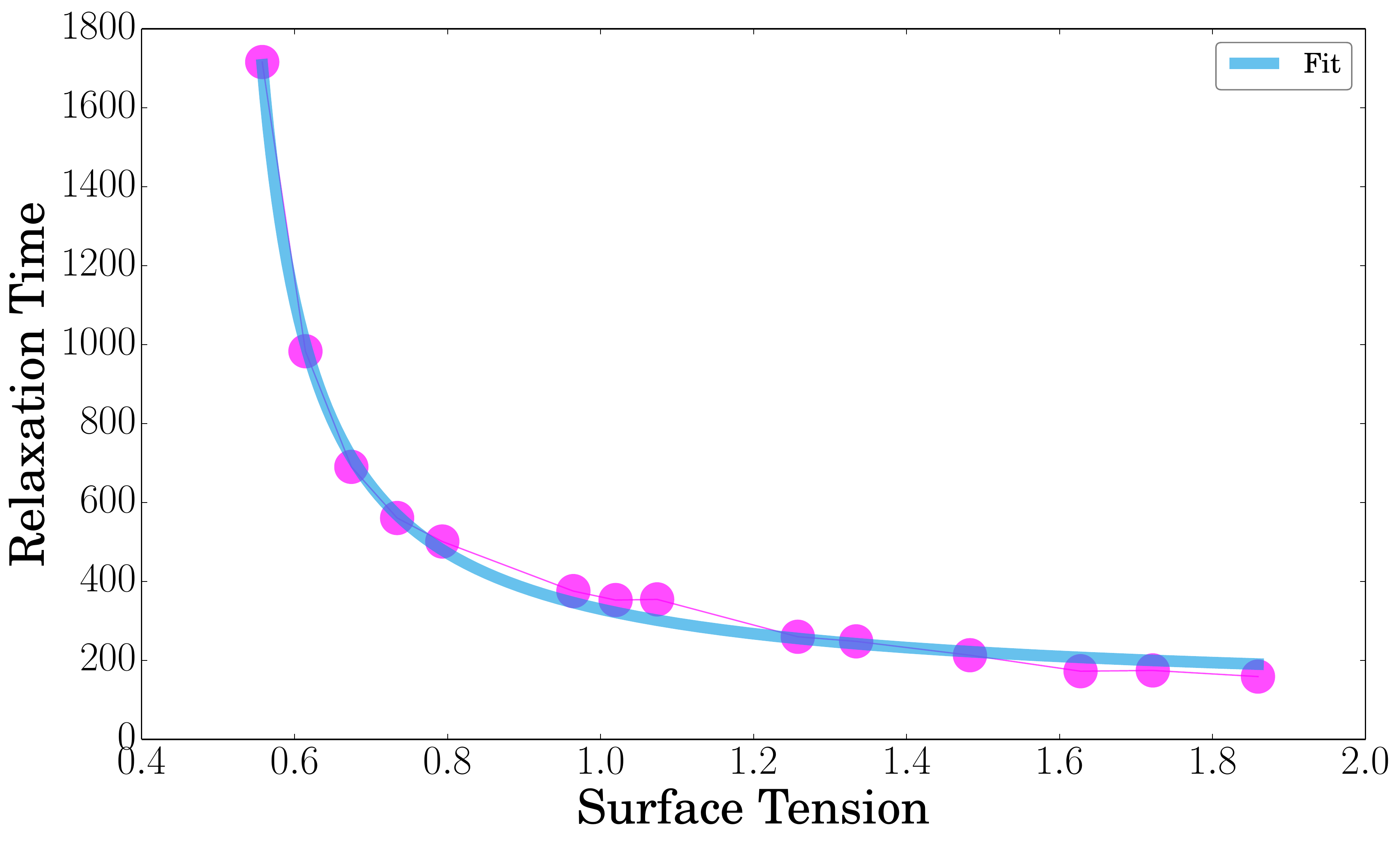} 
\caption[Relaxation time versus surface tension.]{Relaxation time versus surface tension, fit to \autoref{eqn:tauGammaHelfrichFit}. Derived out of the HS model, $\tau(\gamma)$ (\autoref{eqn:tauGammaHelfrich2})  leads to a correct description of the force dependence of the relaxation time. } 
\label{fig:tauVsTension} 
\end{figure}

%▔▔▔▔▔▔▔▔▔▔▔▔▔▔▔▔▔▔▔▔▔▔▔▔▔▔▔▔▔▔▔▔▔▔▔▔▔▔▔▔▔▔▔▔▔▔▔▔▔▔▔▔▔▔▔▔▔▔▔▔▔▔▔▔▔▔▔▔▔▔▔▔▔▔▔▔▔▔▔
\noindent
We are now able to plot $\tau(\gamma)$ ---the relaxation time as a function of surface tension. 
In \autoref{fig:tauVsTension} we plot and fit $\tau (\gamma )$ using \autoref{eqn:tauGammaHelfrichFit}. 
Though the fit extends beyond small $\Delta\gamma$, it does estimate $\eta/\K$ from the asymptote at high tension, which is unchanged in more complicated fitting functions. 

This fit (blue line) gives $\displaystyle \frac{\eta}{\K} \approx   108 \pm 16\Time$, which corresponds to a viscosity
\begin{linenomath*}\begin{equation} \eta \approx  900 \Force\Time/\Length  \approx 3.1\times 10^{-11} \Newton\second/\metre \text{.}    \label{eqn:areaDilatationalViscosityEstimated}\end{equation}\end{linenomath*} 
Interestingly, this viscosity is $\approx 3\times$ the value of $\etaSurface$ (\emph{shear}-surface viscosity) reported by den Otter et al.\cite{den_otter_intermonolayer_2007} for simulated DPPC bilayers. 
(\emph{Dilatational}-surface viscosity $\etaDilatational$ and \emph{shear}-surface viscosity $\etaSurface$ have equivalent dimensions.)
One might expect our $\eta$ to be smaller than that of \cite{den_otter_intermonolayer_2007} since they used longer, two-tailed lipids. 
However for real lipid bilayers, the \emph{dilatational}-surface viscosity $\etaDilatational$ can be two orders of magnitude\cite{dimova_giant_2007, riske_electro-deformation_2005} larger than $\etaSurface$. 
Given this fact, it is actually quite reasonable that our $\eta$ should be larger than \cite{den_otter_intermonolayer_2007}'s $\etaSurface$ as well.

\subsection{Area expansion versus tension} \label{ssec:AlphaVsTension} 
\begin{figure}[h]
\includegraphics[width=0.98\textwidth]{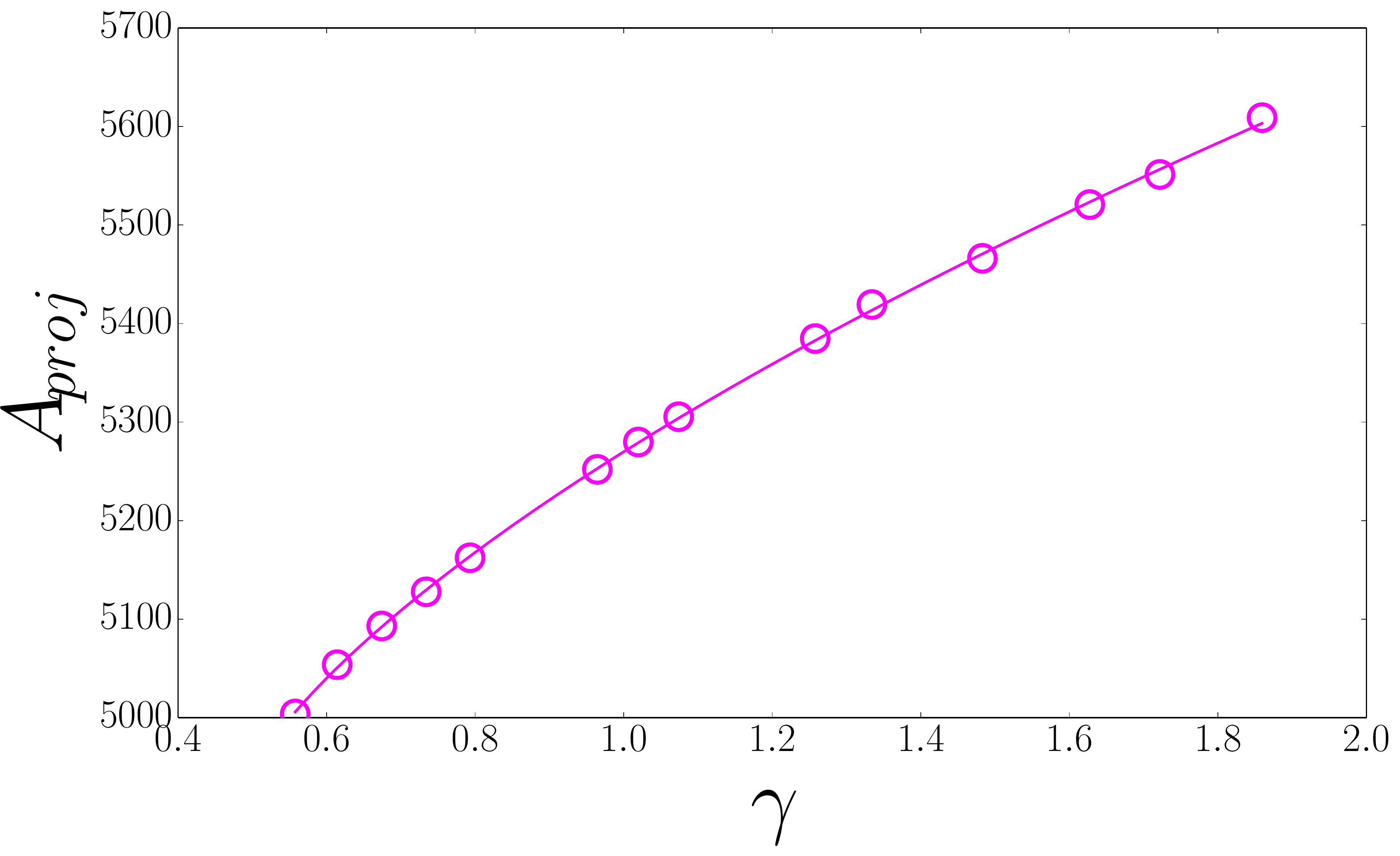} 
\caption[Projected area versus surface tension]{Projected area versus surface tension. The tension ($x$-axis) has been estimated via the work done compressing the vesicle (see \autoref{eqn:tensionWorkDefinition}).} 
\label{fig:AprojHelfrichQUANTITATIVE} 
\end{figure}

%▔▔▔▔▔▔▔▔▔▔▔▔▔▔▔▔▔▔▔▔▔▔▔▔▔▔▔▔▔▔▔▔▔▔▔▔▔▔▔▔▔▔▔▔▔▔▔▔▔▔▔▔▔▔▔▔▔▔▔▔▔▔▔▔▔▔▔▔▔▔▔▔▔▔▔▔▔▔▔
\noindent
\autoref{fig:AprojHelfrichQUANTITATIVE} shows the projected area versus tension. 
The non-linear regime at low tension is characteristic of the entropic behaviour predicted by the HS-model.

%\FloatBarrier
%\FloatBarrier%

\begin{figure}[h]
\includegraphics[width=0.98\textwidth]{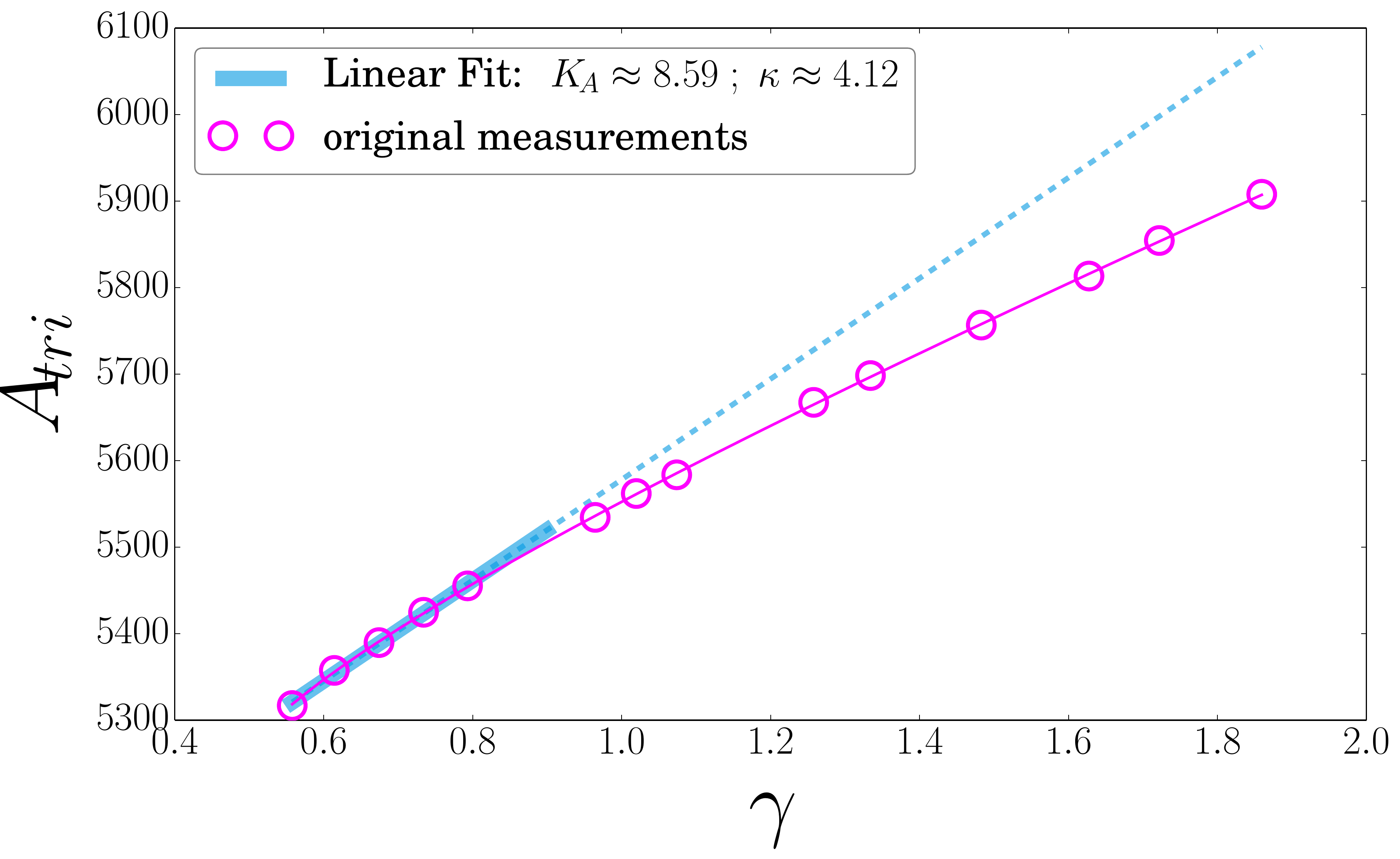} 
\caption[Triangulated area versus surface tension, HS model: estimating $\K$, $\kappa $]{Triangulated area versus surface tension. The tension ($x$-axis) has been estimated via the work done compressing the vesicle (see \autoref{eqn:tensionWorkDefinition}).  $\K$ is estimated using a linear fit to the low tension regime (see \cite{otter_area_2005}). The bending rigidity is $\kappa = \K l^{2}/48$, where $l$ is the bilayer thickness \cite{goetz_mobility_1999}. The $\K$ returned by this fit agrees with that obtained by Bertrand et al.\cite{bertrand_extrusion_2012} (flat bilayer, identical lipids) to within $3\% $. } 
\label{fig:AtriFit} 
\end{figure}

In \autoref{fig:AtriFit} we estimate $\K \approx  8.6 \frac{\!\Energy}{\Length^{2}}$ using a linear fit to the triangulated surface area, in the low-tension regime. 
This value compares well with previous simulations using similar lipids under similar conditions ($\sim 8.8$ \cite{bertrand_extrusion_2012}, $\sim 11.0$ \cite{shkulipa_simulations_2006}, $\sim 13.0$ \cite{otter_area_2005}, $\sim 12.0$-$13.6$ \cite{goetz_computer_1998, goetz_mobility_1999}). Since $\K$ increases with tail length \cite{grafmuller_fusion_2009}, it is reasonable to expect that our value will be at the low end of the spectrum.
Converting our $\K$ into dimensionful units gives $\K \approx 0.1\Newton/\metre$,  which is reasonable when compared with experimental values  ---e.g. AFM indentation of supported bilayers $\K \sim 0.12\Newton/\metre$ \cite{das_nanoscale_2010}, and micropipette aspiration of giant vesicles $\K \sim 0.18\Newton/\metre$ \cite{fa_decrease_2007}, $\sim 0.13$-$0.64\Newton/\metre$ \cite{evans_entropy-driven_1990}. 
A quadratic fit to the triangulated area, like that found in Equation (18) of \cite{otter_area_2005} gives the same $\K$, but requires an additional free parameter.

 %
 %\FloatBarrier
%\FloatBarrier%

\FloatBarrier
\subsection{Vertical compression $\Delta z$} 
\begin{figure}[h]
\includegraphics[width=0.98\textwidth]{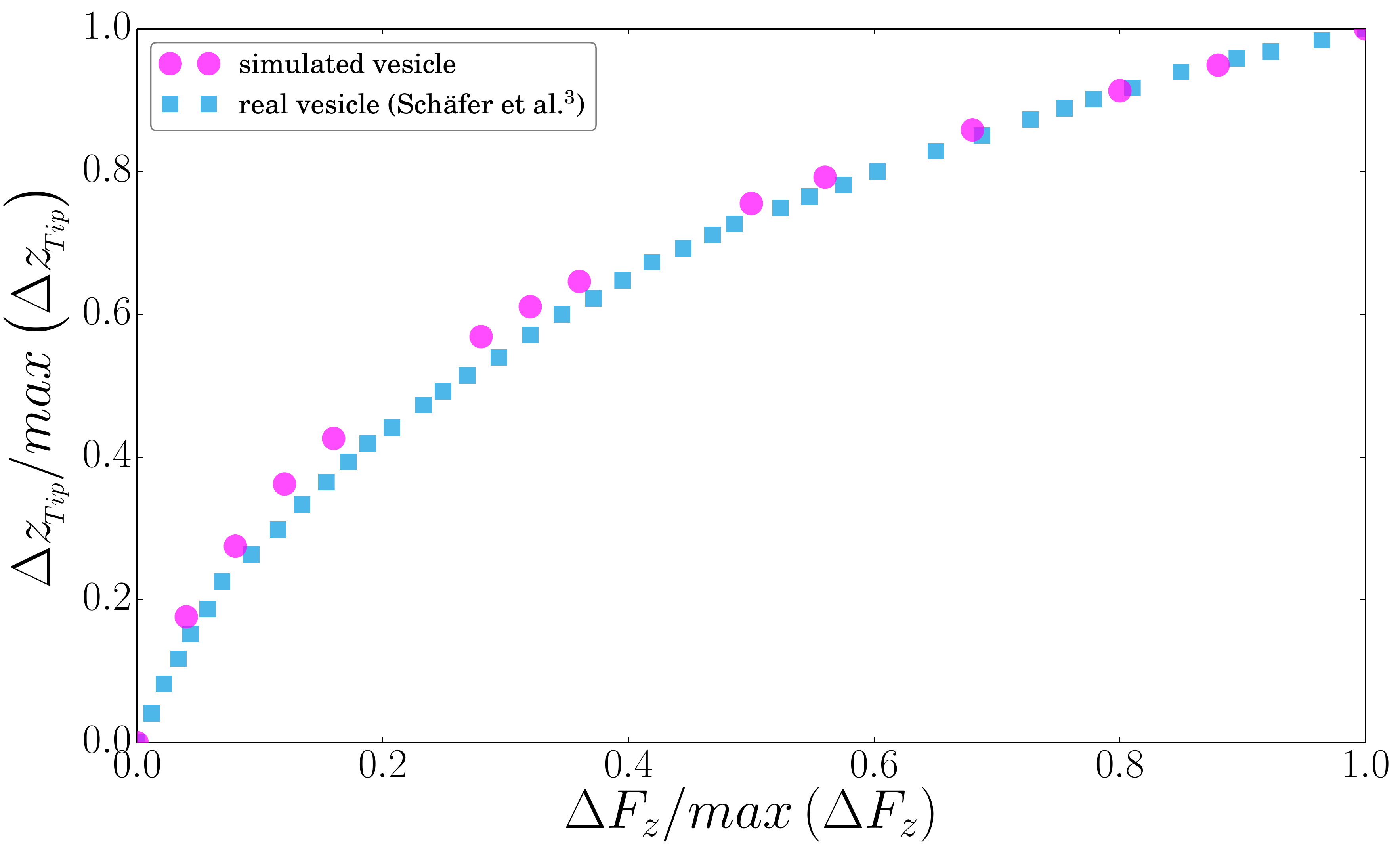} 
\caption[Vertical compression: Relative height variation as a function of applied force]{Vertical compression: Relative height change as a function of applied force. The similarity between  simulation and experimental data (modified from \cite{schafer_mechanical_2013}) is striking.} 
\label{fig:SchaferComparison} 
\end{figure}

%▔▔▔▔▔▔▔▔▔▔▔▔▔▔▔▔▔▔▔▔▔▔▔▔▔▔▔▔▔▔▔▔▔▔▔▔▔▔▔▔▔▔▔▔▔▔▔▔▔▔▔▔▔▔▔▔▔▔▔▔▔▔▔▔▔▔▔▔▔▔▔▔▔▔▔▔▔▔▔
\noindent
In \autoref{fig:SchaferComparison} the vertical compression is scaled as a fraction of the maximum compression which the (respective) vesicle can withstand. 
The scaled GUV data (modified from Sch\"{a}fer et al. \cite{schafer_mechanical_2013}) and our simulation data are very similar; in spite of (i) the immense difference in size and (ii) the fact that our simulations use a  \emph{compressible} fluid  ---experimental buffer solutions are generally incompressible. 
This suggests that it is the physical character of the undulating \emph{membrane} ---rather than the solvent--- that determines the force-compression curve of a fluid-filled vesicle. 
This result is supported by the analysis of Moreno-Flores and Ben\'{\i}tez\cite{moreno-flores_comment_2014}, who found that a vesicle's force-compression curve depends on the properties of its membrane and not on its size. 
%The agreement means that the simulated bilayer vesicle, as a physical system, is practically indistinguishable from a real vesicle in this experiment. 
%
 
Our compression data begins at $\F = 10.0\Force$, whereas the data from Sch\"afer et al.\cite{schafer_mechanical_2013} begins much closer to $\F = 0$. 
For this reason, we make the comparison using $\Delta F_{z} = (F_{z} - F_{z_{0}})$ on the $x$-axis, and $\Delta z = (z - z_{0})$ on the $y$-axis (with $z_{0} = z(F_{z_{0}})$). 

We see from \autoref{fig:SchaferComparison} that $\Delta z$ and $\Delta A$(\autoref{fig:projectedAreaVsForce}) have the same form, however the logarithmic regime in $\Delta z$ is exaggerated compared with that of $\Delta A$. 
This is because initially $\Delta z(F_{z})$ increases more rapidly than $A(F_{z})$.

\section{Discussion} 
%▀▀▀▀▀▀▀▀▀▀▀▀▀▀▀▀▀▀▀▀▀▀▀▀▀▀▀▀▀▀▀▀▀▀▀▀▀▀▀▀▀▀▀▀▀▀▀▀▀▀▀▀▀▀▀▀▀▀▀▀▀▀▀▀▀▀▀▀▀▀▀▀▀▀▀▀▀▀▀▀▀▀▀▀▀▀▀▀▀▀▀▀▀▀▀▀▀▀▀▀▀▀▀▀▀▀▀▀▀▀
\noindent
The relaxation time $\tau$ increases strongly at low tension; essentially $\tau (\gamma )  \sim (\gamma  + \mbox{const})^{-1}$ in this regime. 
This result follows from the Helfrich and Servuss model\cite{helfrich_undulations_1984} (HS model), which describes the steady-state area expansion of bilayer membranes as a function of surface tension. 
%%%SAY MORE ABOUT HOW THIS SETS THE CONTEXT FOR OTHER MEASUREMENTS
The form of $\tau(\gamma)$ given in \autoref{eqn:tauGammaHelfrich} (derived using the HS model) predicts that a vesicle's relaxation time will \emph{depend on its size only at low tension} (see \autoref{eqn:tauGammaBothLimits}). 
Likewise at high tension the relaxation time is predicted to be \emph{independent of the vesicle size}. 

At low tension, flattening of undulations is the dominant form of relaxation (apparent area expansion). 
Vesicle size affects relaxation time by limiting the maximum wavelength and amplitude of these vibrational modes. 
At high tension, direct stretching of the membrane dominates, so the size-effect on the undulations doesn't show up in $\tau$.

%To test these predictions ---that $\tau$ depends on vesicle size at small $\gamma$ and is independent of it at large $\gamma$--- one could measure $\tau$ for a range of vesicle sizes. 
%Real vesicles sized $10\nano\metre \lesssim R\ves  \lesssim 10\micro\metre$ are regularly produced. 
%Simulated vesicles of different sizes might also be used. 

The HS model describes membrane \emph{area expansion} $\alpha (\gamma )$ as the combined effect of flattening entropic undulations and direct stretching. 
The model predicts that our measured area expansion should exhibit curvature in the low tension regime and linearity at high tension, which is what is seen in \autoref{fig:AprojHelfrichQUANTITATIVE}. 

Further analysis in terms of the HS model allowed us to estimate the membrane viscosity, via a curve fit to $\tau (\gamma )$. 
The estimated surface viscosity (\autoref{eqn:areaDilatationalViscosityEstimated}) compares well with that observed for similar bilayers\cite{den_otter_intermonolayer_2007}.

\section{Conclusions} 
%▀▀▀▀▀▀▀▀▀▀▀▀▀▀▀▀▀▀▀▀▀▀▀▀▀▀▀▀▀▀▀▀▀▀▀▀▀▀▀▀▀▀▀▀▀▀▀▀▀▀▀▀▀▀▀▀▀▀▀▀▀▀▀▀▀▀▀▀▀▀▀▀▀▀▀▀▀▀▀▀▀▀▀▀▀▀▀▀▀▀▀▀▀▀▀▀▀▀▀▀▀▀▀▀▀▀▀▀▀▀
\noindent
We report a strong dependence of the relaxation time on applied force (\autoref{fig:tauVsForceNoFit}). %
The effect is greatest at low tension, due to flattening of undulations, but persists until lysis. %
Since undulations have been observed in real vesicles and cells\cite{fricke_variation_1984}, the force dependence should be present in them as well. %
Using the Helfrich and Servuss model\cite{helfrich_undulations_1984} we predict that the effect (in the low force regime) should scale as the surface area (i.e. $\emph{radius}^{2}$) of the membrane. %
Hence the dependence should be strong in real cells and giant vesicles, since their membranes are orders of magnitude larger than our small simulated vesicle.

Relaxation times vary widely in the literature\cite{haase_resiliency_2013, karcher_three-dimensional_2003, desprat_creep_2005, smith_probing_2005, thoumine_time_1997}, and some of this variation may be explained by the results presented above. 
%This may result from varying interactions with the substrate. 
Cells adhere very strongly to some surfaces, and weakly to others depending e.g. on the stiffness of the substrate \cite{yeung_effects_2005, pelham_cell_1997}. %
Strong adhesion suppresses undulations, thereby weakening the force-dependence of $\tau$.  
Experiments are also carried out under different tip conditions\cite{alessandrini2005afm}. % 
We therefore expect the relaxation time to depend strongly on the applied force and on the preexisting tension in the membrane, in short on the experimental setup. %

\section*{Acknowledgments}
%\begin{acknowledgments}
The authors acknowledge support from the Natural Sciences and Engineering Research Council of Canada.
%\end{acknowledgments}

\appendix  %Analysis}  
%▀▀▀▀▀▀▀▀▀▀▀▀▀▀▀▀▀▀▀▀▀▀▀▀▀▀▀▀▀▀▀▀▀▀▀▀▀▀▀▀▀▀▀▀▀▀▀▀▀▀▀▀▀▀▀▀▀▀▀▀▀▀▀▀▀▀▀▀▀▀▀▀▀▀▀▀▀▀▀▀▀▀▀▀▀▀▀▀▀▀▀▀▀▀▀▀▀▀▀▀▀▀▀▀▀▀▀▀▀▀

\noindent
\section{Conversion to dimensionful units} \label{app:UnitsConversion} 
%▔▔▔▔▔▔▔▔▔▔▔▔▔▔▔▔▔▔▔▔▔▔▔▔▔▔▔▔▔▔▔▔▔▔▔▔▔▔▔▔▔▔▔▔▔▔▔▔▔▔▔▔▔▔▔▔▔▔▔▔▔▔▔▔▔▔▔▔▔▔▔▔▔▔▔▔▔▔▔
\noindent
As stated, these unit conversions are presented only as a guideline  ---an approximate scaling of our model to lipid bilayer vesicles. 
The validity of the model is not restricted to vesicles. 
Should this simulation prove relevant to another physical system, another set of unit conversions could of course be invoked. 

The procedure summarized here is based on that used by Goetz and Lipowski\cite{goetz_computer_1998}. 
The energy unit (Lennard-Jones energy) in these simulations is defined $\Energy = \kBoltzmann T$. 
The Lennard-Jones fluid is meant to represent water at SATP, so $\Energy \approx 4\times 10^{-21}\Joule$. 

$\Mass$ is the particle mass. 
That is, every simulated particle is assigned the same mass: $m_{j} \equiv \Mass$. 
A lower-bound on $\Mass$ is the mass of a single $H_{2}O$ molecule ($\approx 3\times 10^{-26}\kilo\gram$). 
Beads making up the tails of the simulated lipids provide an upper bound as they may represent up to six $CH_{2}$ molecules, so that $M \lesssim 14\times 10^{-26}\kilo\gram$. 

A lower bound for $\Length$ (the Lennard-Jones length) is the average separation between two solvent molecules. 
For water, this is $\approx  0.31\nano\metre$. 
If a lipid tail bead represents at most six $CH_{2}$ groups, then the maximum distance between these beads along the lipid chain is six carbon-carbon bond lengths ($\approx 0.9 \nano\metre$). 
 
The unit of simulation time is $\Time = \sqrt{\Mass\Length^{2}/\Energy}$, and the force unit is $\Force = \Energy/\Length$. 
Plugging in the above conversions yields  $\Time \approx 2.8\pico\second$, and $\Force \approx 6.6\pico\Newton$.

\noindent
\section{Surface area of bilayer ---triangulated surface} \label{ssec:TriangulatedSurface} 
%▔▔▔▔▔▔▔▔▔▔▔▔▔▔▔▔▔▔▔▔▔▔▔▔▔▔▔▔▔▔▔▔▔▔▔▔▔▔▔▔▔▔▔▔▔▔▔▔▔▔▔▔▔▔▔▔▔▔▔▔▔▔▔▔▔▔▔▔▔▔▔▔▔▔▔▔▔▔▔
\noindent
The relaxation process was observed via the triangulated surface area, which is a direct measurement of the surface area of the vesicle. 
Triangulation: the bilayer's inner and outer leaflets are each approximated as tessellated surfaces, composed of triangles whose vertices are located at the lipid heads. 
Adding up the surface area of all the triangles composing the tessellated surface gives its total surface area ---which we call the ``triangulated area'' of the membrane. 
The triangulated area is much closer to the true area of the membrane, rather than its apparent (i.e. `projected') area. 

%%%INCLUDE GRAPHIC OF TESSELLATED VESICLE
%%%GIVE MORE DETAILS ABOUT TRIANGULATION ALGORITHM

The derivation of $\tau(\gamma)$ in \autoref{derHelfrich} was done in terms of projected area, but relaxation times were obtained by fitting the \emph{triangulated} (rather than projected) area. 
This complication does not harm the analysis. 
Notice that the direct area expansion term in \autoref{eqn:alphaHelfrich} is $\frac{\gamma}{\K}$, so direct stretching of the membrane is nonzero even at low tension (when flattening of undulations dominates the relaxation). %
\autoref{eqn:alphaHelfrich}\cite{helfrich_undulations_1984} treats undulation flattening ($\alphaEntropic$) and direct stretching ($\alphaDirect$) like two springs in series\footnote{Squeezing the vesicle increases its internal pressure, which increases membrane tension ---`pulling on the springs'. } which relax simultaneously ---pulling on either spring stretches both. 

In short, we assume that there is one relaxation time, the time required for the system to reach steady state. %
Using a wave expansion of the undulations, Helfrich and Servuss calculated the apparent area of a membrane. %
So this is what was used in our theory. %
In numerical simulations triangulation methods can accurately estimate the surface area of the membrane. %
The relaxations times were obtained from the time variation of the triangulated area. %

\noindent
\section{Surface area of bilayer ---apparent surface} \label{app:ApparentSurface} 
\begin{figure}[h]
\includegraphics[width=0.98\textwidth]{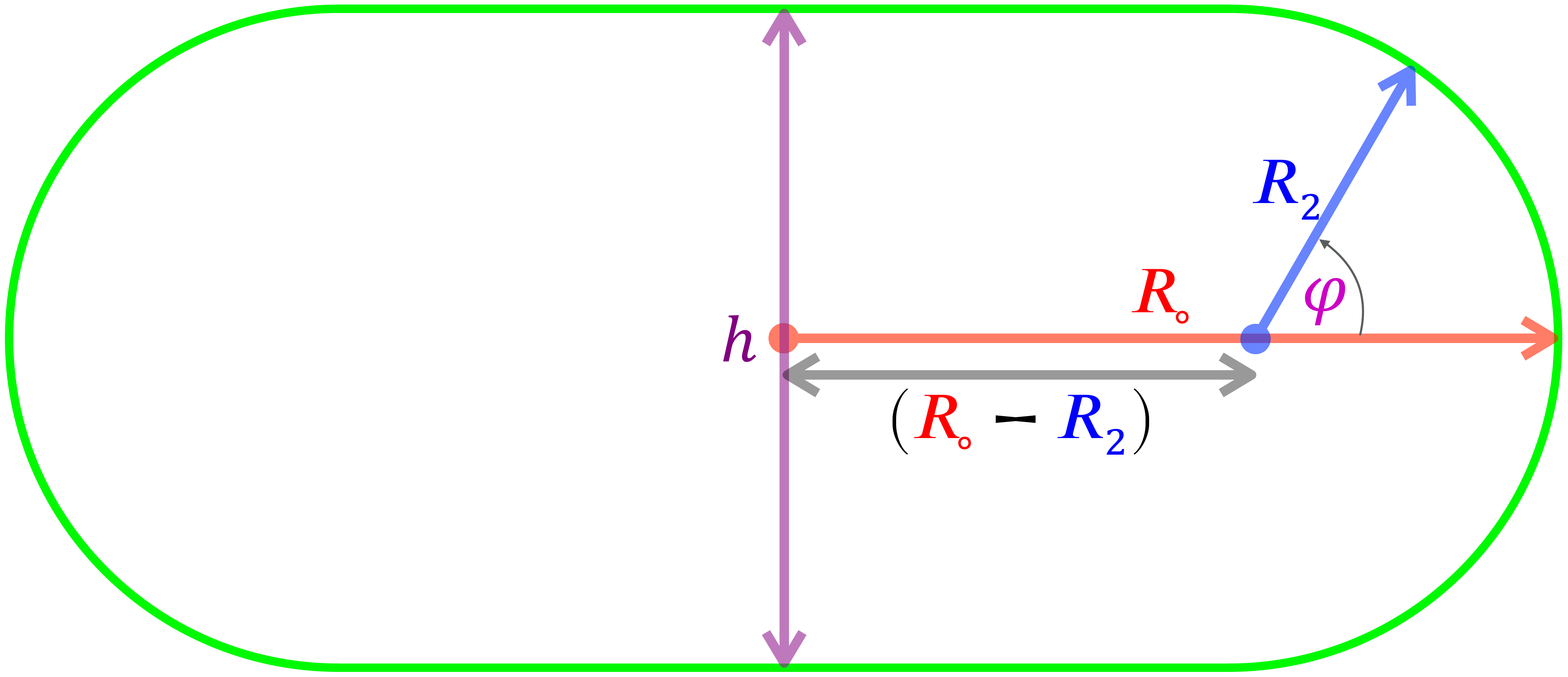} 
\caption[Parametrization of compressed vesicle]{Parametrization of a compressed vesicle (cf. \autoref{fig:vesicleProfile}). In the context of projected area and membrane undulations, this profile delineates the apparent surface of the vesicle. The shape of the compressed vesicle is well approximated by a ``filled torus'' ---a doughnut without a hole.% (start with a cylinder and let its outer edge bulge out into a circular arc)
This approximation of the true surface is the same as was used by \cite{schafer_mechanical_2013, yoneda_tension_1964} and detailed in \cite{evans_mechanics_1980}. The variable names have been chosen to match those of \cite{schafer_mechanical_2013} for ease of comparison. Coordinates: In this figure, $h$ lies along the $z$-axis. $\phi$ lies in the $xy$-plane. $\varphi = \arctan(z/R_{2})$.} 
\label{fig:vesicleParametrization} 
\end{figure}

\begin{figure}[h]
\includegraphics[width=0.98\textwidth]{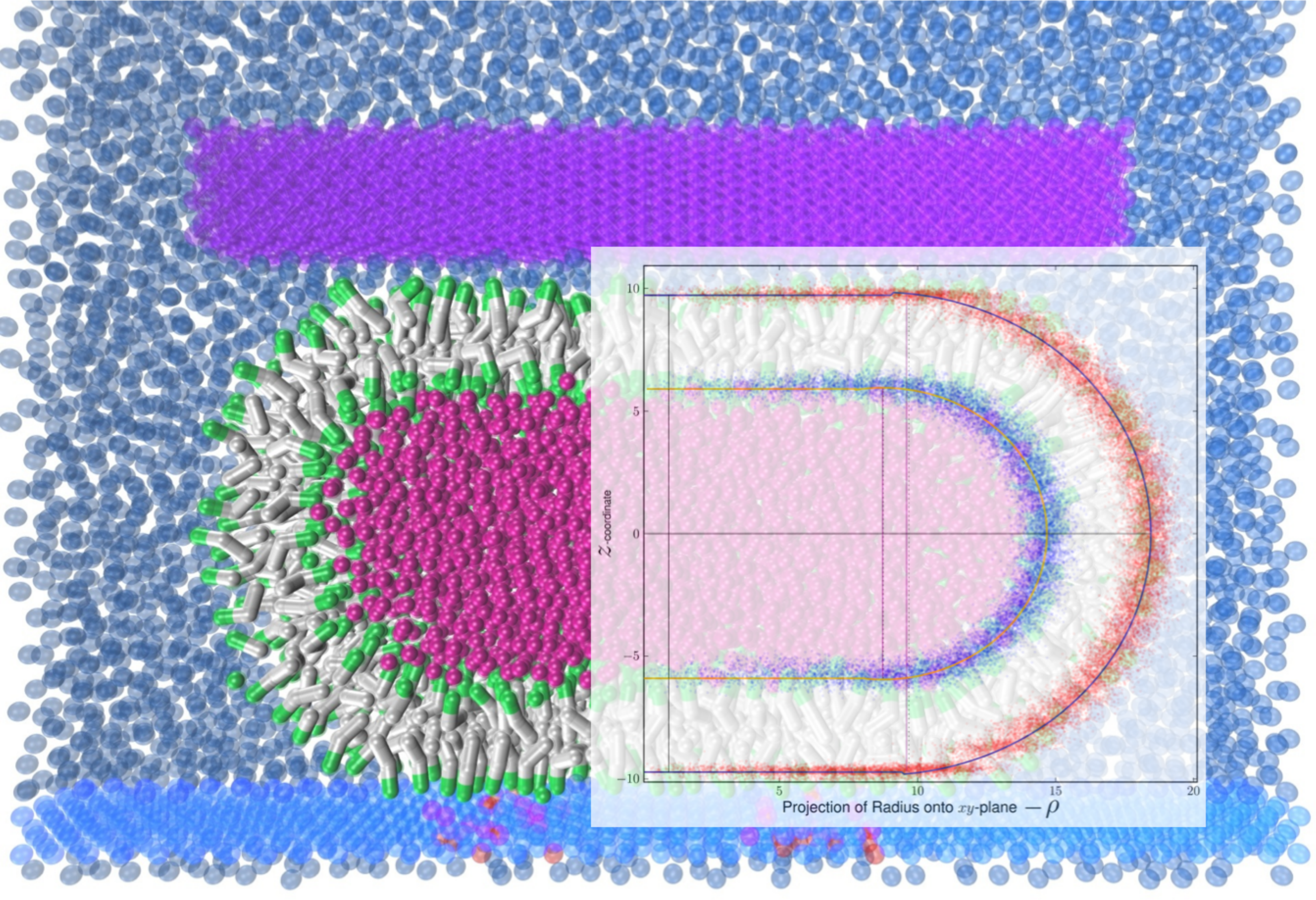} 
\caption[Profile of simulated vesicle]{Vesicle profile and the corresponding fit which measures the parameters $h$, $R_{0}$ and $R_{2}$ shown in \autoref{fig:vesicleParametrization}. } 
\label{fig:vesicleProfile} 
\end{figure}

%▔▔▔▔▔▔▔▔▔▔▔▔▔▔▔▔▔▔▔▔▔▔▔▔▔▔▔▔▔▔▔▔▔▔▔▔▔▔▔▔▔▔▔▔▔▔▔▔▔▔▔▔▔▔▔▔▔▔▔▔▔▔▔▔▔▔▔▔▔▔▔▔▔▔▔▔▔▔▔
\noindent
The apparent area expansion of vesicles is described by the HS model\cite{helfrich_undulations_1984} (see \autoref{eqn:alphaHelfrich}). 
To measure the apparent area, the vesicle shape is parameterized (\autoref{fig:vesicleParametrization}) and curve-fit (\autoref{fig:vesicleProfile}). 

The free surface of the compressed vesicle (curved region in Figures \ref{fig:vesicleParametrization} and \ref{fig:vesicleProfile}) is described by the position vector
\begin{linenomath*}\begin{equation}  \vec{r}(\phi , z) = \rho (z)\hat{e}_{\rho }(\phi ) + z\hat{e}_{z} \qquad \substack{ 0 \leq \phi < 2\pi \\\\ -\frac{h}{2} \leq z \leq \frac{h}{2} } \label{eqn:freeSurface}\end{equation}\end{linenomath*} 
in cylindrical coordinates $(\rho , \phi , z)$ with 
\begin{linenomath*}\begin{equation}  \rho (z) = (R_{0} -R_{2}) + \sqrt{R_{2}^{2} - z^{2}} \text{,}  \label{eqn:rho(z)}\end{equation}\end{linenomath*} 
 and
\begin{linenomath*}\begin{equation} \hat{e}_{\rho }(\phi ) = \cos\phi \hat{e}_{x} + \sin\phi \hat{e}_{y}\text{.}  \label{eqn:eRho}\end{equation}\end{linenomath*} 
Curve-fitting the free surface allows us to parametrize the vesicle's \emph{entire} apparent surface ---from which we calculate the apparent area.

\noindent
\section{Revised Helfrich-Servuss Model} \label{app:RevisedHSmodel} 
%▔▔▔▔▔▔▔▔▔▔▔▔▔▔▔▔▔▔▔▔▔▔▔▔▔▔▔▔▔▔▔▔▔▔▔▔▔▔▔▔▔▔▔▔▔▔▔▔▔▔▔▔▔▔▔▔▔▔▔▔▔▔▔▔▔▔▔▔▔▔▔▔▔▔▔▔▔▔▔
As demonstrated in Figure 2 of  Mell et al.\cite{mell_fluctuation_2015}, the undulation spectrum $P = P\HS(\ell, \gamma)$ that was used by Helfrich and Servuss\cite{helfrich_undulations_1984} to derive the HS model departs from experimental fluctuation spectra at high wavenumber $\ell$.

The relevance of this to the model that we presented for the vesicle's relaxation time $\tau(\gamma)$ is as follows: 
We model the compressed vesicle's relaxation time as a function of tension $\gamma$, in terms of an effective stiffness $K(\gamma)$ and viscosity $\eta$:
 \begin{linenomath*}
\begin{equation} \tau \sim \frac{\eta}{K} \sim K^{-1}(\gamma)\text{,} \label{eqn:tauDefK}\end{equation}
 \end{linenomath*}
with $K^{-1} = \frac{\partial \alpha}{\partial \gamma}$.
$\alpha$, which denotes the relative change in a membrane's apparent area due to the competing effects of entropic undulations and surface tension, is modelled using the expression derived by Helfrich and Servuss \cite{helfrich_undulations_1984}.
The HS model contains two terms
 \begin{linenomath*}
\begin{equation}  \alpha(\gamma) = \alphaEntropic(\gamma) + \alphaDirect(\gamma)  \text{.} \label{eqn:alphaHelfrichMnemonicAppendix}\end{equation}
 \end{linenomath*}
The first term gives the fraction of membrane area `absorbed' by undulations, reducing the `apparent area' of the membrane, and the second term accounts for direct stretching of the membrane by surface tension.

Here we are concerned only with the former, `entropic' term. 
To obtain it, Helfrich and Servuss integrate over the spectrum of undulations 
 \begin{linenomath*}
\begin{equation} \alphaEntropic(\gamma) \propto \int_{\ell_{\text{min}}}^{{\ell_{\text{max}}}}  \ell^{3} P\HS(\ell, \gamma)d\ell \text{,}\label{eqn:HSintegral}\end{equation}
 \end{linenomath*}
where equation (9b) of Helfrich and Servuss (1984) gives the spectrum as
\begin{linenomath*}
\begin{equation}  P\HS(\ell, \gamma) \approx \frac{\kBoltzmann  T}{\gamma \ell^{2} + \kappa \ell^{4}} \text{.} \label{eqn:HSspectrum}\end{equation}
\end{linenomath*}
Since in our model $\tau(\gamma) \sim \frac{\partial \alpha}{\partial \gamma} = \frac{\partial}{\partial \gamma}\left(\alphaEntropic + \alphaDirect \right) $, the undulation spectrum $P(\ell, \gamma)$ directly affects our model of $\tau(\gamma)$ through $\alphaEntropic$.
\\

Having arrived at the relevant point, we ask: Do the experimental spectra $P\Mell(\ell, \gamma)$ in \cite{mell_fluctuation_2015} differ from the approximation used by Helfrich and Servuss in such a way as to alter our model of $\tau(\gamma)$? We claim that the answer is ``no'': 
Equation (16a) in \cite{mell_fluctuation_2015} gives the bimodal spectrum
\begin{linenomath*}
\begin{equation} P\Mell(\ell, \gamma) \approx \frac{\kBoltzmann  T}{\gamma \ell^{2} + \kappa \ell^{4}} + \left(\frac{12R}{h\K}\right)\frac{\kBoltzmann T}{\ell^{2} + \frac{1}{2}hR\ell^{4}}  \label{eqn:PMell}\end{equation}
\end{linenomath*}
which was fit to their observations. 
The departure of the observed spectra from the HS model can be expressed by writing 
\begin{linenomath*}
\begin{equation}  P\Mell(\ell, \gamma) = P\HS(\ell, \gamma) + f(\ell)\text{.}\label{eqn:MellSpectrum}\end{equation}
\end{linenomath*}
That is to say, the experimental spectra differ from the HS model by a term which \emph{does not depend on $\gamma$}. 
 Hence that while the spectrum $P\Mell(\ell, \gamma)$ adds another entropic term $\alphaEntropic'$ to the HS model 
\begin{linenomath*}
\begin{equation} \alphaEntropic(\gamma) \propto \int_{\ell_{\text{min}}}^{{\ell_{\text{max}}}} \ell^{3} P\HS(\ell, \gamma)d\ell + \underbrace{\int_{\ell_{\text{min}}}^{{\ell_{\text{max}}}} \ell^{3} f(\ell) d\ell}_{\alphaEntropic' \propto}\text{,}\label{eqn:HSintegralMell}\end{equation}
\end{linenomath*}
this difference is moot when we take its $\gamma$-deriviative  as outlined above, to model $\tau(\gamma)$.
\\

While the additional entropic term does not affect our model for the relaxation time, it does add some additional detail to the HS model. 
We obtain a `Revised HS model' by recapitulating Helfrich and Servuss' derivation, this time using $P\Mell$.
Their derivation assumes that the local inclination $\phi(\vec{r})$ is small  ($\tan\phi << 1 $) even at high wavenumber.
At large $\ell$, the amplitude of the undulations decays more slowly in $P\Mell$ than in $P\HS$ (see Figure 2 in \cite{mell_fluctuation_2015}), so we must check that the small $\phi$ assumption still holds.

For a given mode $u_{\ell}(\vec{r})$ with amplitude $u_{\ell}$, we have $\tan(\phi_{\ell}) = |\nabla{u_{\ell}(\vec{r})}| \lesssim  \ell u_{\ell}$.
Since $P(\ell) \propto \langle u_{\ell}^{2} \rangle$, we write
\begin{linenomath*}
\begin{equation} \frac{\langle\tan^{2}{{\phi_{\ell,}}\Mell}\rangle}{\langle\tan^{2}{{\phi_{\ell,}}\HS}\rangle} \lesssim \frac{P\Mell}{P\HS}  \approx 1 + \left(\frac{12 R}{h\K }\right)\frac{\gamma + \kappa\ell^{2}}{1 + \frac{1}{2}hR\ell^{2}} \text{.}\label{eqn:PHelfrichOverPmell}\end{equation}
\end{linenomath*}
So if the small-$\phi$ approximation is valid for $P\HS$ then it is valid for $P\Mell$ as well, provided the term on the right hand side of Equation \ref{eqn:PHelfrichOverPmell} is not too large.
To obtain an upper bound on this term, we let the wavenumber go to infinity
\begin{linenomath*}
\begin{equation}  \limit{\ell}{\infty}\left(\frac{12 R}{h\K }\right)\frac{\gamma + \kappa\ell^{2}}{1 + \frac{1}{2}hR\ell^{2}}  =  \frac{24\kappa }{h^{2} \K} \implies \frac{P\Mell}{P\HS} \leq \left(1 +  \frac{24\kappa }{h^{2} \K}\right)\text{.} \label{eqn:UpperLimitOnPMelloverPHSTerm0}\end{equation}
\end{linenomath*} 
Using $h_{\text{\!effective}} \approx 2\text{nm}$, $R \lesssim 20 \mu\text{m}$,  $\K \approx 0.1 \text{N}/\text{m}$, $\gamma \lesssim 1 \mu\text{N}/\text{m}$, and $\kappa \approx 20\kBoltzmann T$,  we have
\begin{linenomath*}
\begin{equation} \frac{P\Mell}{P\HS}  \lesssim 6 \text{,} \label{eqn:UpperLimitOnPMelloverPHSTerm1}\end{equation}
\end{linenomath*}
with
\begin{linenomath*}
\begin{equation} \limit{\ell}{0}\frac{P\Mell}{P\HS} =  \left(1 + \frac{12 R\gamma }{h\K }\right) \approx 2.2\text{.} \label{eqn:LowerLimitOnPMelloverPHSTerm0}\end{equation}
\end{linenomath*}

The two spectra have the same order of magnitude, and therefore the weighting of modes used by Helfrich and Servuss\cite{helfrich_undulations_1984} 
 to integrate the area absorption over the spectrum of undulations remains valid:
 \begin{linenomath*}
\begin{equation}  (\Delta A)_{\ell} \propto -\ell^{2}P(\ell, \gamma) \text{.}\label{eqn:HelfrichServussWeighting}\end{equation}
\end{linenomath*} 

Using $P\Mell$ (\autoref{eqn:PMell}),  the entropic term in $\alpha(\gamma)$ becomes 
\begin{linenomath*}
\begin{equation}  \alphaEntropic(\gamma) = \frac{1}{4 \pi} \int_{\ell_{\text{min}}}^{{\ell_{\text{max}}}} \ell^{3}P\Mell(\ell, \gamma) d\ell \approx  \frac{\kBoltzmann T}{4 \pi} \int_{\ell_{\text{min}}}^{{\ell_{\text{max}}}} \left[\frac{\ell}{\gamma  + \kappa \ell^{2}} + \left(\frac{12R}{h\K}\right)\frac{\ell}{1 + \frac{1}{2}hR\ell^{2}}\right] d\ell\text{,}\label{eqn:RevisedHelfrich0}\end{equation}
\end{linenomath*} 

\begin{linenomath*}
\begin{equation} =\frac{\kBoltzmann T}{4 \pi}  \left[\frac{1}{2 \kappa }\ln{\left(\frac{\gamma  + \kappa \ell^{2}_{\text{min}}}{\gamma  + \kappa \ell^{2}_{\text{max}}}\right)} + \left(\frac{12}{h^{2}\K}\right)\ln{\left(\frac{1 + \frac{1}{2}hR\ell^{2}_{\text{min}}}{1 + \frac{1}{2}hR\ell^{2}_{\text{max}}}\right)}\right] \text{.}\label{eqn:RevisedHelfrich2}\end{equation}
\end{linenomath*} 

\noindent
Writing the cutoffs $\ell^{2}_{\text{min}} = \frac{\zeta}{A}, \ell^{2}_{\text{max}} = \frac{\zeta}{a}$, we arrive at a `revised HS model':
\begin{linenomath*}
\begin{equation}  \alpha(\gamma) \approx \underbrace{\frac{\kBoltzmann T}{8\pi \kappa }\ln{\left(\frac{\frac{\zeta }{A} + \frac{\gamma }{\kappa }}{\frac{\zeta }{a} + \frac{\gamma }{\kappa }}\right)} + \frac{3\kBoltzmann T}{\pi h^{2} \K }\ln{\left(\frac{1 + \frac{hR\zeta }{2A}}{1 + \frac{hR\zeta }{2a}}\right)}}_{\emph{entropic}} + \underbrace{\frac{\gamma }{\K}}_{\emph{direct}} \label{eqn:RevisedHelfrich1}\end{equation}
\end{linenomath*} 
(c.f. \autoref{eqn:alphaHelfrich}). The middle term, which results from the high wavenumber correction contained in $P\Mell$, shifts $\alpha(\gamma)$ by a constant but does not alter its tension-dependence.

\bibliography{Bibliography}
\bibliographystyle{apsrev4-1}

 \end{document}